%% file: belike2.tex
\documentclass[usenatbib]{mn2e}

\usepackage[varg]{txfonts}
\usepackage{bm}
\usepackage{graphicx}
\usepackage{latexsym}
\usepackage{amssymb}
\usepackage{subfigure}
\usepackage{supertabular}
\usepackage{natbib,twoopt}

\usepackage{color}

\usepackage[breaklinks=true]{hyperref} %% to avoid \citeads line fills
\bibpunct{(}{)}{;}{a}{}{,} %% natbib format for A&A and ApJ
\makeatletter
\newcommandtwoopt{\citeads}[3][][]{\href{http://adsabs.harvard.edu/abs/#3}%
{\def\hyper@linkstart##1##2{}%
\let\hyper@linkend\@empty\citealp[#1][#2]{#3}}}
\newcommandtwoopt{\citepads}[3][][]{\href{http://adsabs.harvard.edu/abs/#3}%
{\def\hyper@linkstart##1##2{}%
\let\hyper@linkend\@empty\citep[#1][#2]{#3}}}
\newcommandtwoopt{\citetads}[3][][]{\href{http://adsabs.harvard.edu/abs/#3}%
{\def\hyper@linkstart##1##2{}%
\let\hyper@linkend\@empty\citet[#1][#2]{#3}}}
\newcommandtwoopt{\citeyearads}[3][][]%
{\href{http://adsabs.harvard.edu/abs/#3}
{\def\hyper@linkstart##1##2{}%
\let\hyper@linkend\@empty\citeyear[#1][#2]{#3}}}
\makeatother

\def\re{\mathrm{e}}

\def\Kelvin{\,\mathrm{K}}
\def\Ry{\,\mathrm{Ry}}

\title[Validity of the ICFT $R$-matrix method]
      {Validity of the ICFT $R$-matrix method:
       Be-like Al~$^{9+}$ a case study}
\author[L. F. Menchero et al]{
        L. Fern\'andez-Menchero$^{1}$\thanks{
        E-mail: luis.fernandez-menchero@strath.ac.uk (LFM); \newline
        badnell@phys.strath.ac.uk (NRB); \newline
        gd232@cam.ac.uk (GDZ)}, 
        G. Del~Zanna$^{2}$ and N. R. Badnell$^{1}$ \\
        $^{1}$Department of Physics, University of Strathclyde. 
              Glasgow G4 0NG, United Kingdom \\
        $^{2}$Department of Applied Mathematics and Theoretical Physics, 
              University of Cambridge, Cambridge CB3 0WA, United Kingdom
       }

\begin{document}

\date{Accepted tomorrow. Received \today; 
      in original form yesterday}

\pagerange{\pageref{firstpage}--\pageref{lastpage}} \pubyear{2015}

\maketitle

\label{firstpage}

\begin{abstract}

We have carried-out 98-level configuration-interaction / close-coupling (CI/CC) 
intermediate coupling frame transformation (ICFT) and Breit--Pauli $R$-matrix 
calculations for the electron-impact excitation of Be-like Al~$^{9+}$. The close
agreement that we find between the two sets of effective collision strengths
demonstrates the continued robustness of the ICFT method. 
On the other hand, a comparison of this data with previous 238-level 
CI/CC ICFT effective collision strengths shows that the results for
excitation up to $n=4$ levels are systematically and increasingly 
underestimated over a wide range of temperatures by $R$-matrix calculations 
whose close-coupling  expansion extends only to $n=4$ (98-levels). 

Thus, we find to be false a recent conjecture that the ICFT approach may 
not be completely robust. The conjecture was based upon a comparison of 
98-level CI/CC Dirac $R$-matrix effective collision strengths for Al~$^{9+}$
with those from the 238-level CI/CC ICFT $R$-matrix calculations. The disagreement found
recently is due to a lack of convergence of the close-coupling expansion in the 
98-level CI/CC Dirac work.
The earlier 238-level CI/CC ICFT work has a superior target
to  the 98-level CI/CC Dirac one and provides more accurate atomic data.

Similar considerations need to be made for other Be-like ions and for other sequences.

\end{abstract}

\begin{keywords}
Atomic data --  Techniques: spectroscopic
\end{keywords}

\section{Introduction}
\label{sec:introduction}

Electron-impact excitation is the dominant process for populating the
radiating states of ions whose emission lines form the basis for the
spectroscopic diagnostic modelling of non-equilibrium astrophysical
and laboratory plasmas.
As such, a great deal of effort over many years has gone in to calculating
a large amount of collision data, which is incorporated into databases
and modelling suites such as
CHIANTI\footnote{\url{http://www.chiantidatabase.org}} 
and OPEN-ADAS\footnote{\url{http://open.adas.ac.uk}}.
The pre-eminent methodology used is the $R$-matrix one.
However, there are many variations on this theme.

The intermediate coupling frame transformation (ICFT) $R$-matrix method is 
an approximation to the Breit--Pauli $R$-matrix method (BPRM) which neglects 
the spin--orbit interaction between the colliding electron and the ion.
Thus, it is physically well motivated. 
There is a good deal of literature which verifies the accuracy of the ICFT 
approach: the original comparisons of the results of ICFT and Breit--Pauli 
$R$-matrix calculations by \citet{griffin1998,griffin1999} 
on $\mathrm{Mg}$-like ions and \citet{badnell1999a} on $\mathrm{Ni}^{4+}$; 
more recent ones by \citet{liang2010a} comparing ICFT $R$-matrix and Dirac 
Atomic $R$-matrix Code (DARC) calculations for $\mathrm{Ne}$-like 
$\mathrm{Fe}^{16+}$ and $\mathrm{Kr}^{26+}$, \citet{liang2009b} on ICFT with 
Breit--Pauli and DARC for $\mathrm{Na}$-like $\mathrm{Fe}^{15+}$; 
and most recently \citet{badnell2014} compared the results of ICFT, BPRM and 
DARC calculations for $\mathrm{Fe}^{2+}$. 
The differences observed between ICFT and other $R$-matrix results are all 
well within the uncertainties to be expected due to the use of different 
configuration interaction and close-coupling expansions and resonance 
resolution. 
Indeed, the \citet{badnell2014} work on $\mathrm{Fe}^{2+}$ used identical 
atomic structures and close-coupling expansions for the ICFT and Breit--Pauli 
$R$-matrix calculations and found excellent agreement, better than $5\%$. 
Their DARC calculations out of necessity used a somewhat different atomic 
structure, and the low-level structure of $\mathrm{Fe}^{2+}$ is challenging, 
but still gave agreement to within $\sim 20\%$ at $10^4\Kelvin$.

Extensive calculations have been carried-out applying the ICFT $R$-matrix 
method to whole isoelectronic sequences (for elements, typically, up 
to $\mathrm{Zn}$), the most recent ones being 
$\mathrm{Mg}$-like \citep{fernandez-menchero2014b}, 
$\mathrm{Be}$-like \citep{fernandez-menchero2014a} and 
$\mathrm{B}$-like \citep{liang2012} --- this
last paper contains references to earlier sequences.

Thus it is both surprising and of great concern to find a work which counters this 
trend: \citet{aggarwal2015} made a comparison of the results of a 98-level 
DARC calculation on $\mathrm{Be}$-like Al~$^{9+}$
\citep{aggarwal2014b} with a 238-level ICFT $R$-matrix one 
\citep{fernandez-menchero2014a}. \citet{aggarwal2015} found that the 
effective collision strengths obtained  from the 238-level 
ICFT $R$-matrix  calculation were significantly larger than the DARC ones in 
many instances and they suggested that the results of 
\citet{fernandez-menchero2014a} were less reliable, querying the ICFT method,
resonance resolution and its high energy / temperature behaviour.

In addition, \citet{aggarwal2015} alighted on the recent paper 
by \citet{storey2014b} which reported a problem in the outer-region ICFT 
calculation of $\mathrm{O}^{2+}$, when compared to a full Breit--Pauli 
calculation, and suggested that this could be the main cause of the 
discrepancies for Al~$^{9+}$. 
However, \citet{storey2014b} noted that such an issue only arises when 
resonance effective quantum numbers become small. 
The problem is peculiar to low-charge ions such as $\mathrm{O}^{2+}$ and 
unusually small $R$-matrix box sizes.
\citet{storey2014b} focused on providing a solution to their problem at hand.
We note that the problem does not arise in the first place 
if the $R$-matrix box size is increased (beyond its default in their case) to 
encompass a spectroscopic $n=3$ orbital, say. 
This is why the issue had not arisen before: all previous calculations 
(including \cite{fernandez-menchero2014a}) used larger, often much larger,
box sizes. 
The problem noted by \cite{storey2014b} is not relevant in general.
\citet{aggarwal2015} noted a similar trend for other ions of the Be-like 
sequence $\mathrm{Cl}^{13+}$, $\mathrm{K}^{15+}$, $\mathrm{Ge}^{28+}$ 
\citep{aggarwal2014a} and $\mathrm{Ti}^{18+}$~\citep{aggarwal2012a}.

Where does that leave us with regard to the discrepancies noted by 
\citet{aggarwal2015}? The concern of \citet{aggarwal2015} was the 
disagreement between $R$-matrix calculations of (apparent) comparable complexity. 
However, the works of \citet{aggarwal2014b} and 
\citet{fernandez-menchero2014a} are not of comparable complexity. 
We show here that the much larger close-coupling expansion used by 
\citep{fernandez-menchero2014a} (238 vs 98 levels) gives rise to a systematic
enhancement of effective collision strengths over a wide range of
temperatures, which increases as one excites higher-and-higher levels. 

In addition, we analyze the uncertainty in the effective collision strengths 
due to the incompleteness of the configuration interaction expansion, the
validity of the ICFT vs Breit--Pauli method, viz. the neglect of the spin--orbit
interaction of the colliding electron, and the effect of resonance 
resolution and position on low temperature effective collision strengths.

The paper is organized as follows. 
In section \ref{sec:methods} we describe the methodology we used for the 
different calculations we have performed. 
In section \ref{sec:structure} we discuss the atomic structure 
of Al~$^{9+}$ and present results for energies, line strengths and
infinite energy plane wave Born collision strengths.
In section \ref{sec:upsilons} we compare and contrast effective collision 
strengths. 
In section \ref{sec:conclusions} we present our main conclusions.
Atomic units are used unless otherwise specified.

\section{Methodology}
\label{sec:methods}

In the following sections we compare the results of the 238-level 
configuration-interaction / close-coupling (CI/CC) ICFT $R$-matrix calculation 
by \citet{fernandez-menchero2014a} for Al~$^{9+}$ with the results of
new ICFT and Breit--Pauli 98-level CI/CC $R$-matrix calculations --- 
the latter being the same sized CI \& CC expansions that were used 
by~\cite{aggarwal2012a,aggarwal2014a,aggarwal2014b}. 
Where possible (meaningful), our new 98-level CC calculations follow the same 
prescription as \citet{fernandez-menchero2014a} e.g. with respect to angular 
momentum and energy specification.

The target description uses the {\sc autostructure} program 
\citep{badnell2011b}.
\citet{fernandez-menchero2014a} included all of the configurations 
$\mathrm{1s^2\,\{2s^2,2s2p,2p^2\}}$ and $\mathrm{1s^2\,\{2s,2p\}}\,nl$  with 
$n=3-7$ and $l=\mathrm{s,p,d,f,g}$  for $n=3-5$ and $l=\mathrm{s,p,d}$ 
for $n=6,7$, which makes a total of 238 levels. 
In the present work, we restrict ourselves to $n=4$ 
(with $l=\mathrm{s,p,d,f}$) which gives a total of 98 levels. 
The Thomas--Fermi potential scaling parameters, $\lambda_{nl}$, for the atomic 
structure calculation with the CI basis set of 98 levels are given in 
Table \ref{tab:lambda}, those for the CI calculation with 238 levels are 
given in~\citet{fernandez-menchero2014a}.
The $\lambda_{nl}$ parameters for the 98-level CI calculation were obtained in 
the same way as the ones for the 238-level calculation 
\citep{fernandez-menchero2014a}: by minimizing the equally-weighted sum of 
all $LS$-coupling term energies.

\begin{table}
   \caption{\label{tab:lambda} Scaling parameters $(\lambda_{nl})$
             used in the 98-level CI structure calculation.}
\begin{center}
\begin{tabular}{cc}
   \hline 
   Orbital & $\lambda_{nl}$ \\
   \hline
   1s & 1.57175 \\
   2s & 1.29675 \\
   2p & 1.17081 \\
   3s & 1.29004 \\
   3p & 1.15958 \\
   3d & 1.29791 \\
   4s & 1.29797 \\
   4p & 1.15413 \\
   4d & 1.30212 \\
   4f & 1.50916 \\
   \hline
\end{tabular}                      
\end{center}
\end{table}

For the collision calculations we use the inner-region $R$-matrix programs 
of \citet{hummer1993,berrington1995} and the outer-region {\sc stgf} program
of \citet{berrington1987,badnell1999c} plus the ICFT one 
of \citet{griffin1998}.
We will compare the results of ICFT and Breit--Pauli $R$-matrix methods to treat 
relativistic effects in the {\it scattering problem}. 
Both can use the exact same Breit--Pauli {\it atomic structure}. 
This is important as it enables us to isolate differences due solely to the 
differing treatment of relativistic effects in the scattering.
We note that \citet{berrington2005} have shown that the Breit--Pauli 
$R$-matrix method gives essentially the same results as the Dirac one 
for $Z\lesssim 30$.

The ICFT $R$-matrix method first carries-out an $LS$-coupling close-coupling 
calculation for the CI target described above and it includes the mass-velocity 
and Darwin one-body relativistic operators.
The $LS$-coupling reactance $K$-matrices are first recoupled to $jK$-coupling
and then transformed to intermediate coupling using the term coupling 
coefficients \citep{hummer1993} for the Breit--Pauli target.
This imposes the exact same atomic structure on the ICFT $K$-matrices as a
Breit--Pauli one which uses the same CI expansion and radial orbitals.
In particular, the fine-structure levels within a term are non-degenerate
in the final scattering calculation.
This method is the one which was used by~\citet{fernandez-menchero2014a}.

The second method which we use is the Breit--Pauli (BP) one.
In this method the close-coupling expansion is made in intermediate
coupling as well, in addition to the target CI, i.e. the one-body effective 
(nuclear plus Blume \& Watson) spin--orbit operator is included in the
$(N+1)$-electron Hamiltonian.
The Breit--Pauli formalism increases considerably the size of the Hamiltonian
matrix, which makes it impractical to use this method for 238~CC levels.

The essential physical difference between the ICFT and Breit-Pauli $R$-matrix 
methods is the neglect by the former of the effective spin--orbit interaction
between the colliding electron and the ion.
The practical benefit of the ICFT method over the Breit--Pauli one is the 
diagonalization of much smaller $(N+1)$-electron Hamiltonian matrices and a 
much smaller set of coupled scattering equations to be solved in the 
outer-region by {\sc stgf}.

\section{Structure}
\label{sec:structure}

\input{energies.tex}

Table \ref{tab:energies} compares our energies for the first 98 levels of 
Al~$^{9+}$ calculated with the 98- and 238-level CI targets.
These energies are compared also with the observed ones tabulated in the
NIST\footnote{\url{http://physics.nist.gov}} database, which were taken
from work of \citet{martin1979}.
There are ten levels in Table \ref{tab:energies} which do not follow the  same
order in both structure calculations: from level index 82 to 84 and from 88 to 94.
We use the order of the 98-level CI calculation to index levels for comparison purposes.
Energies calculated with both CI expansions have differences smaller than 
$0.5\%$ with the observed ones for most of the levels, and differences
of around $2\%$ for the low lying singlet states.
In general, the differences are similar to those found 
by \citet{aggarwal2014b}.
In the original paper from \citet{martin1979} there are several gaps in the
level energies, and some of them are labeled as ``inaccurate'', so we refer 
the reader to the original work to avoid hasty conclusions.

It is difficult to relate differences in energies directly to differences
in collision data. 
However, differences in oscillator or line strengths ($S$) and infinite energy 
plane-wave Born collision strengths ($\Omega^{\rm PWB}_{\infty}$), 
essentially non-dipole electric multipole line strengths, can be so related. 
\citet{burgess1992} show how infinite energy / temperature and
ordinary / effective collision strengths ($\Omega_{\infty}/\Upsilon_{\infty}$) 
from any scattering calculation, including an $R$-matrix one, are directly 
related to these quantities, viz.
\begin{equation}
   \Upsilon_{\infty}\ =\ \Omega_{\infty}\ =\ \Omega^{\rm PWB}_{\infty}
\end{equation}
for non-dipole allowed transitions, while for electric dipole ones
\begin{equation}
   \Omega_{\infty}\ =\ 
   \lim_{E \rightarrow \infty} \frac{4S}{3}{\ln \left( \frac{E}{\Delta E} + 
   {\rm e} \right)} \,,
\end{equation}
and
\begin{equation}
   \Upsilon_{\infty}\ =\ 
   \lim_{T \rightarrow \infty} \frac{4S}{3}{\ln \left( \frac{kT}{\Delta E} + 
   {\rm e} \right)} \,,
\end{equation}
where $\Delta E$ is the excitation energy for the transition.

In practice, we find that changes in the line strength, $S$, 
or $\Omega^{\rm PWB}_{\infty}$ between two different atomic structures change 
not only the infinite energy values but also the (background) ordinary 
collision strength  correspondingly over a wide range of collision energies, 
and hence the effective collision strength over a wide range of temperatures, 
unless dominated by resonances. 
A $20\%$ change, say, in $S$ or  $\Omega^{\rm PWB}_{\infty}$ provides a very  
realistic measure of the resulting change in effective collision strengths.
Thus, care must be taken when attempting to deduce anything about scattering 
methods from differences in the collision data without reference to 
differences in the underlying atomic structure.

\begin{figure}
\centering
   \includegraphics[width=\columnwidth,clip]{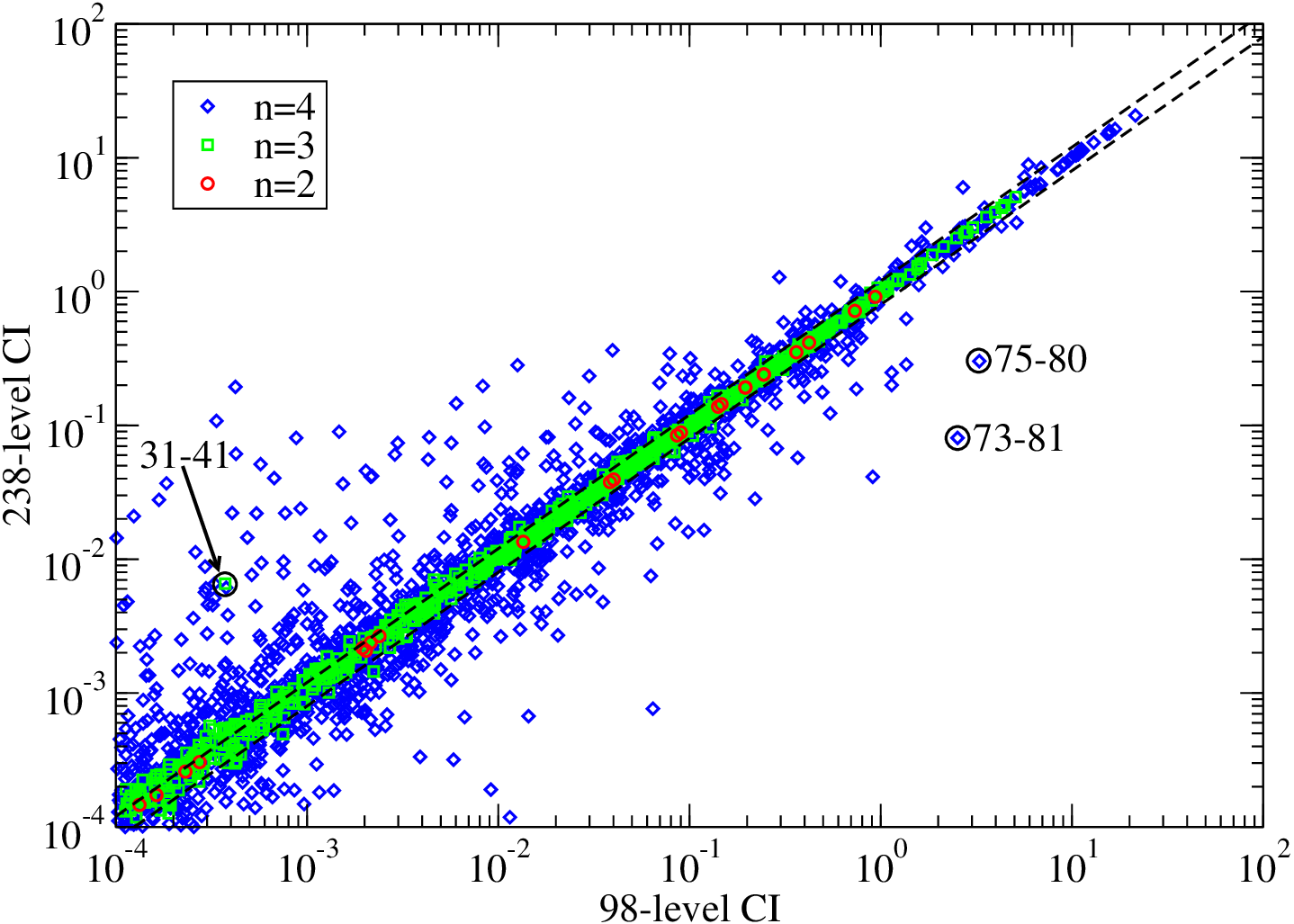}
   \caption{A comparison of $y_{\infty}$ (see eq.~\ref{eq:yinfty}) for the 
      98- vs 238-level CI atomic structures for transitions amongst the 
      98 lowest common levels of Al~$^{9+}$.
      $\circ$: transitions with upper level with $n=2$; 
      $\square$: transitions with upper level with $n=3$; 
      $\diamond$: transitions with upper level with $n=4$;
      dashed lines: $20\%$ fractional difference.
      }
   \label{fig:upsvups4i7i_inf}
\end{figure}

To illustrate, we show in Fig.~\ref{fig:upsvups4i7i_inf} a comparison of the 
reduced quantity $y_{\infty}$, given by
\begin{equation}
   y_{\infty}\ =\ 
    \Omega^{\rm PWB}_{\infty}\quad\mbox{ or} \quad \frac{4S}{3} \,,
\label{eq:yinfty}
\end{equation}
for the two atomic structure calculations which we consider 
(which we label 98-level CI and 238-level CI).
We show points for a total of 4035 transitions resulting from the common 98 
levels.
These split into 1466 dipole transitions and 2569 Born-allowed transitions, 
while there are an additional 718 forbidden transitions that are not 
represented.
We highlight by colour and symbol transitions to upper levels 
with $n=2,3,4$.
We see that transitions up to $n=2$ are very well converged. 
Only one transition differs by more than $10\%$; it is the very 
weak ($\sim 10^{-8}$) Born transition $6-9$:
$\mathrm{2p^2\,^3P_0 - 2p^2\,^1D_2}$ (off the scale).
There is good convergence for transitions up to $n=3$, with 233 out of 801 
transitions which differ by more than $20\%$ but mostly for the weaker 
transitions.
However, there is much more spread for transitions up to $n=4$, 
1834 of 3202 transitions differ by more than $20\%$.
This plot illustrates that the $n=4$ levels of the 98-level CI structure are 
not so well converged with respect to the CI expansion. 
We expect that their representation by the 238-level CI expansion to be much 
better converged.

We circle several transitions for comment. 
The transition $31-41$: $\mathrm{2p3p\,^3P_1 - 2p3d\,^3P_2^{o}}$
is a dipole one, its line strength changes by a factor of 18 between the 
98- and 238-level structures.
Note also the transitions
$75-80$: $\mathrm{2p4d\,^1D_2^{o} - 2p4f\,^1F_3}$ and
$73-81$: $\mathrm{2p4d\,^3F_2^{o} - 2p4f\,^3F_3}$.
Both transitions are dipole allowed and quite strong and yet differ by about 
a factor of 20 and 30 between the two structures. 
These levels lie towards the upper end of the 98-level CI expansion.
These differences highlight the need for as accurate an atomic description as 
possible to obtain the best target for accurate collision data and the need 
to exercise extreme caution  when making comparisons of collision data based on 
different atomic structures.

\section{Collisions}
\label{sec:upsilons}

In this section, we carry-out a series of comparisons of effective collision 
strengths for Al~$^{9+}$ at the temperature of peak abundance for an electron
collisional plasma \citep{bryans2006}.
We then look at issues relating in particular to low temperature (e.g. photoionized) 
plasmas and much higher temperature plasmas (e.g. solar flares).

\subsection{Peak abundance temperature}
\label{subsec:peak}

\begin{figure}
\centering
   \includegraphics[width=\columnwidth,clip]{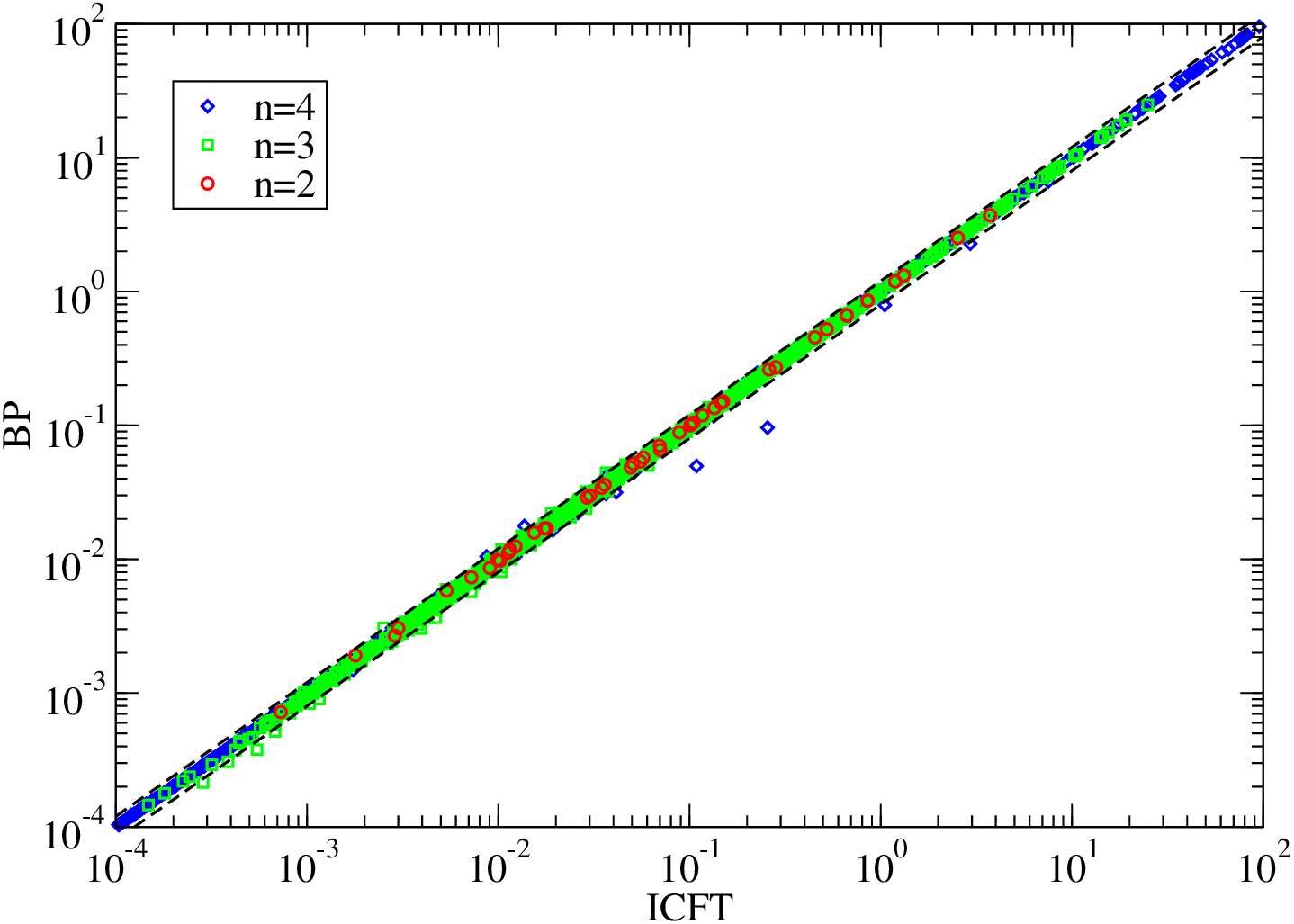}
   \caption{A comparison of effective collision strengths from 98-level CI/CC 
      ICFT versus Breit--Pauli $R$-matrix calculations for all inelastic transitions 
      amongst the 98 levels of Al~$^{9+}$
      at $T_\mathrm{e} = 10^6 \Kelvin$.
      $\circ$: transitions with upper level  $n=2$; 
      $\square$: transitions with upper level  $n=3$; 
      $\diamond$: transitions with upper level  $n=4$;
      dashed lines:  $20\%$ fractional difference.
      }
   \label{fig:upsvups4b4i_1e6K}
\end{figure}

\begin{figure}
\centering
   \includegraphics[width=\columnwidth,clip]{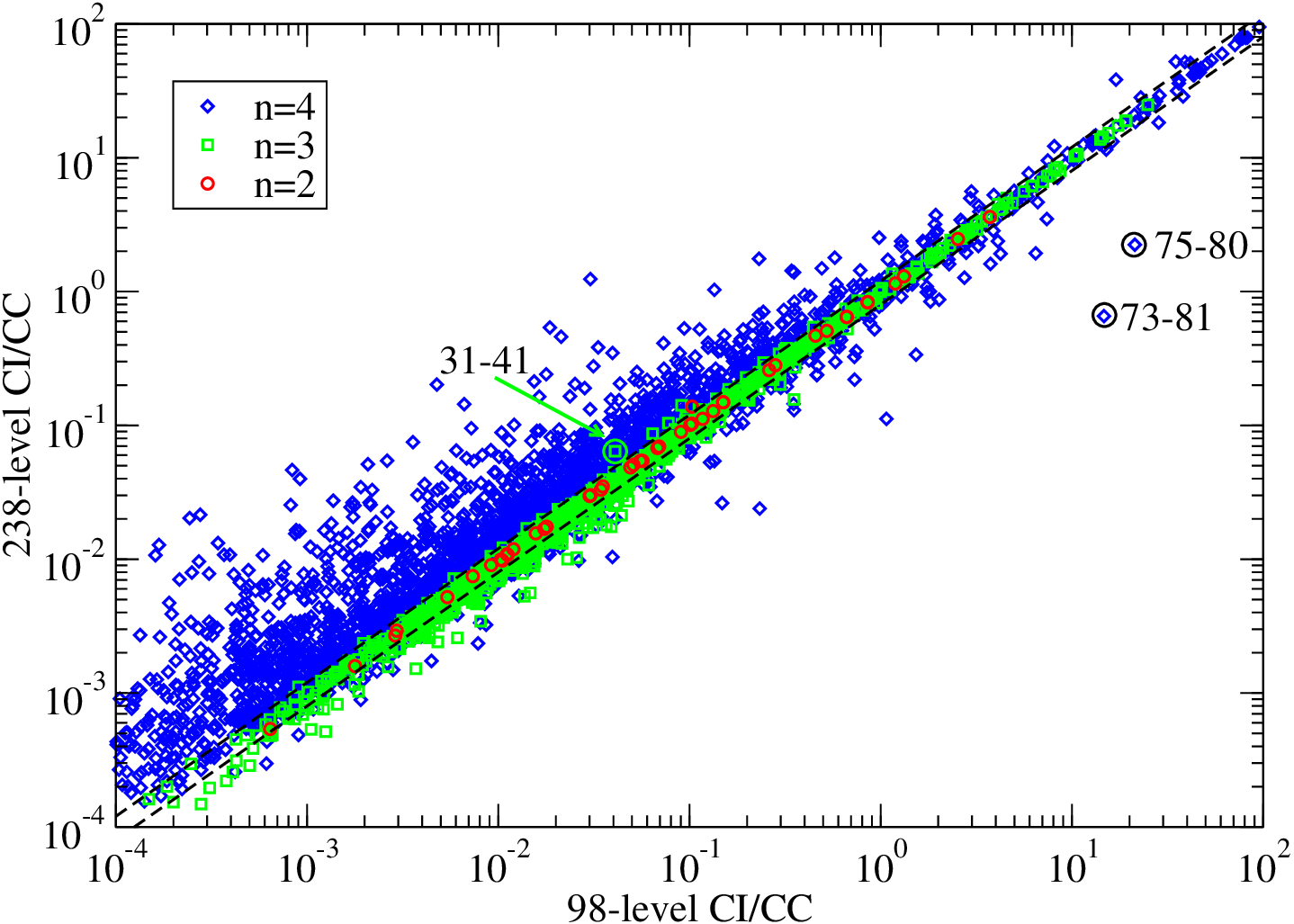}
   \caption{A comparison of effective collision strengths from 98-
       vs 238-level CI/CC ICFT $R$-matrix calculations for all inelastic 
      transitions amongst the 98 lowest common levels 
      of Al~$^{9+}$ at $T_\mathrm{e} = 10^6 \Kelvin$.
      $\circ$: transitions with upper level  $n=2$; 
      $\square$: transitions with upper level  $n=3$; 
      $\diamond$: transitions with upper level  $n=4$;
      dashed lines:  $20\%$ fractional difference.
      }
   \label{fig:upsvups4i7i_1e6K}
\end{figure}

In Fig.~\ref{fig:upsvups4b4i_1e6K} we compare our 98-level CC ICFT and 
Breit--Pauli $R$-matrix Maxwellian effective collision strengths, $\Upsilon$, 
for all inelastic transitions in Al~$^{9+}$ at the temperature of
peak abundance, $10^6\Kelvin$, in an electron collisional plasma.
We recall that both calculations use the exact same (98-level CI) structure.
We note excellent agreement between the two scattering methods, as 
illustrated by the points lying on the diagonal, only 82 points out of 4753 
have a difference larger than $10\%$. 
This is to be contrasted with the spread of points shown in 
Fig.~\ref{fig:upsvups4i7i_inf} which results from our two different atomic 
structures.

Only two points differ from the diagonal more than $50\%$, at around
$\Upsilon=10^{-1}$, they correspond to the transitions
$67-75$: $\mathrm{2p4p\,^3D_2 - 2p4d\,^1D_2^{o}}$ and
$83-98$: $\mathrm{2p4f\,^3F_2 - 2p4d\,^1P_1^{o}}$.
Both are dipole allowed through spin--orbit mixing only.
It is likely that including the spin--orbit interaction in the collision
calculation causes the difference, providing additional mixing via the 
$(N+1)$-electron Hamiltonian diagonalization.

Next, in Fig.~\ref{fig:upsvups4i7i_1e6K} we compare our 98- versus 
238-level CI/CC ICFT effective collision strengths as a whole. 
In contrast to Fig.~\ref{fig:upsvups4b4i_1e6K}, we see a much wider spread of 
points, i.e. the agreement between different scattering methods is much (much) 
better than that obtained using different configuration interaction basis 
sets.
Like the comparison of line strengths and $\Omega^{\rm PWB}_{\infty}$ shown in 
Fig.~\ref{fig:upsvups4i7i_inf}, we note the excellent and very good agreement 
between the two sets of results for transitions up to $n=2$ and $n=3$, 
respectively, while there is a much wider spread in the comparison for 
transitions up to $n=4$. 
However, unlike the atomic structure comparison, there is a systematic shift above 
the diagonal. 
The 238-level CC results are systematically larger than the 98-level ones. 
Thus, we conclude that this is not due to the differences in atomic structure, 
which were  evenly distributed above 
and below the diagonal, rather that this is a measure of the 
lack of convergence of the close-coupling expansion in the 98-level CC 
calculation for the transitions involving $n=4$ levels.
This mirrors the lack of convergence of the 98-level configuration 
interaction expansion for $n=4$ levels. 
We note that there are still more than one hundred levels which lie above
the $n=4$ levels in the 238-level CI/CC case.

Specifically, one half of the transitions in Fig.~\ref{fig:upsvups4i7i_1e6K} 
differ by more than $30\%$. 
The number of transitions differing by more than $20\%$ (lines indicated in the plot)
correspond to 1 for $n=2$ (out of 45), 256 for $n=3$ (out of 990) and 2331 
for $n=4$ (out of 3718).
Note also that the 3 transitions which were circled in 
Fig.~\ref{fig:upsvups4i7i_inf} are again circled  in 
Fig.~\ref{fig:upsvups4i7i_1e6K}. 
The strong transitions $75-80$ and $73-81$ do illustrate how outliers in the 
atomic structure comparison (factor 20 and 30 difference) show-up as outliers 
in the collision comparison (factor 10 and 20 difference).
For the weaker transition $31-41$ the resonances in the collision calculation 
``dampen'' the difference in atomic structure, it being ``just'' a factor 
of $1.6$ now.
In Table \ref{tab:density} we give the exact number of transitions which have
a difference $\delta=|\Upsilon_{98}-\Upsilon_{238}|/\Upsilon_{238}$
larger than a given percentage for the $y_{\infty}$ for the
98- vs 238-level CI atomic structures as well as the $\Upsilon$ at $T=10^6 \Kelvin$ 
for the 98-level CC ICFT vs Breit--Pauli and the 98- vs 238-level CC ICFT comparisons.

\begin{figure}
\centering
   \includegraphics[width=\columnwidth,clip]{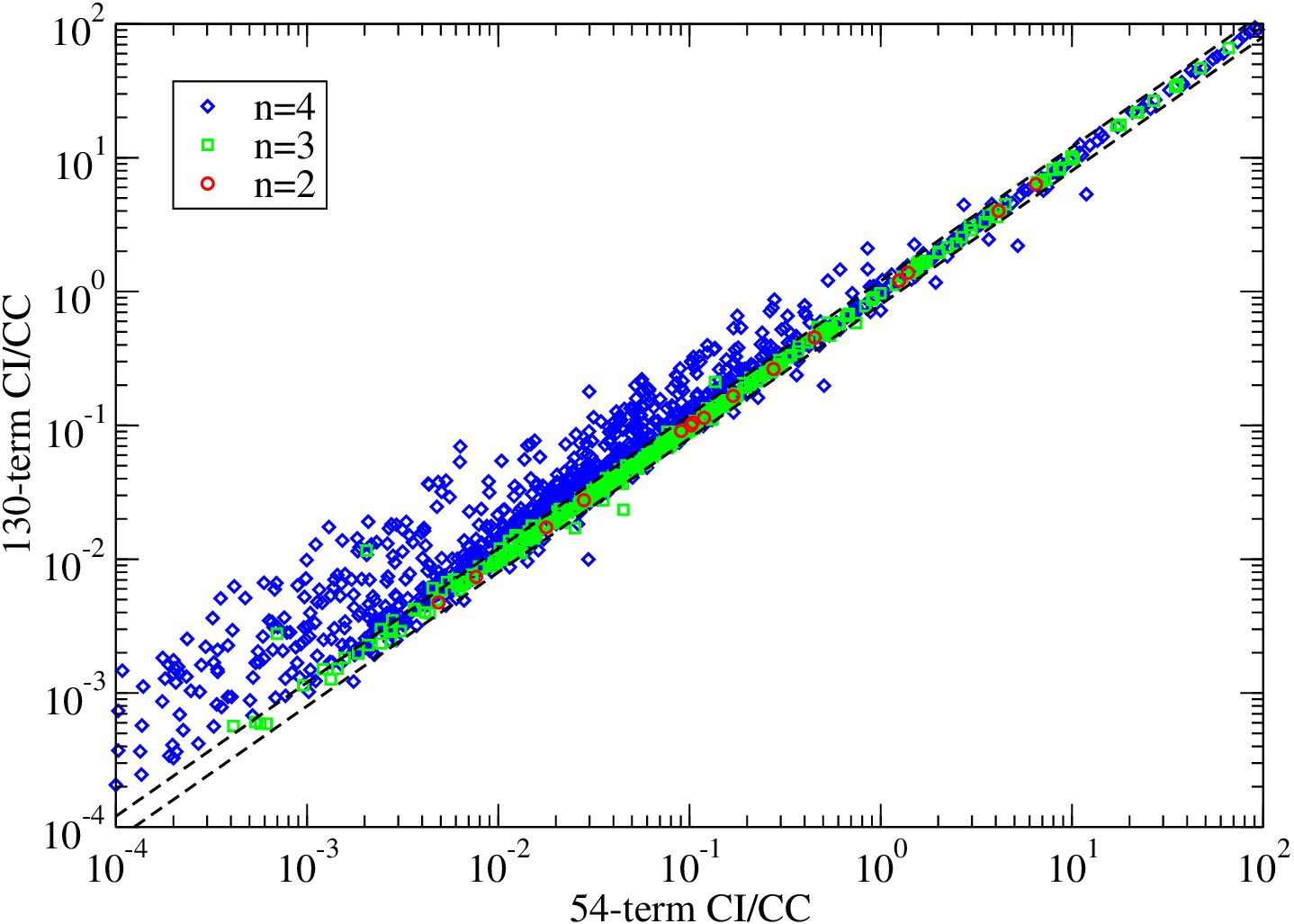}
   \caption{A comparison of effective collision strengths from 54-
      vs 130-term CI/CC $LS$-coupling $R$-matrix calculations for all 
      inelastic transitions amongst the 54 lowest common terms 
      of Al~$^{9+}$ at $T_{\mathrm{e}} = 10^6 \Kelvin$.
      $\circ$: transitions with upper term  $n=2$; 
      $\square$: transitions with upper term  $n=3$; 
      $\diamond$: transitions with upper term  $n=4$;
      dashed lines:  $20\%$ fractional difference.
      }
   \label{fig:upsvups4l7l_1e6K}
\end{figure}

We can confirm further that the systematic increase of the effective collision
strengths to $n=4$ in the 238-level CC calculation over those of the 
98-level CC calculation is due to the lack of convergence of the 
close-coupling expansion in the latter.
Like the convergence of the CI expansion, the convergence of the 
close-coupling expansion is essentially independent of the coupling scheme, 
i.e. the specific $R$-matrix method used, be it LS, ICFT, Breit--Pauli or DARC. 
We illustrate this in Fig.~\ref{fig:upsvups4l7l_1e6K} where we make a similar 
comparison of 54- vs 130-term CC $LS$-coupling $R$-matrix effective collision 
strengths.
We see the same systematic increase for transitions to $n=4$ as in the 
comparison of ICFT $R$-matrix effective collision strengths.

\begin{table}
\begin{minipage}{\columnwidth}
   \caption{\label{tab:density} Number of transitions in Figs 
      \ref{fig:upsvups4i7i_inf} -- \ref{fig:upsvups4i7i_1e6K} which 
      differ by more than a certain relative error
      $\delta=|\Upsilon_{98}-\Upsilon_{238}|/\Upsilon_{238}$ 
     (or $\Upsilon \rightarrow y_\infty$), as a percentage.}
\begin{center}
\begin{tabular}{cccc}
   \hline 
     & Fig.~\ref{fig:upsvups4i7i_inf}: $y_{\infty}$ & 
     Fig.~\ref{fig:upsvups4b4i_1e6K}: 98 CI/CC $\Upsilon$ & 
     Fig.~\ref{fig:upsvups4i7i_1e6K}: ICFT $\Upsilon$ \\
     Rel. error ($\%$)& 98 vs 238 CI & BP vs ICFT & 98 vs 238 CI/CC \\
   \hline
      1  &  3778  &  1336  &  4579  \\
      2  &  3600  &   803  &  4400  \\
      3  &  3416  &   500  &  4243  \\
      4  &  3266  &   350  &  4077  \\
      5  &  3141  &   260  &  3928  \\
      6  &  3022  &   206  &  3798  \\
      7  &  2914  &   158  &  3676  \\
      8  &  2804  &   127  &  3569  \\
      9  &  2722  &   106  &  3460  \\
     10  &  2644  &    82  &  3357  \\
     20  &  2068  &    22  &  2582  \\
     30  &  1643  &     9  &  2090  \\
     40  &  1356  &     4  &  1725  \\
     50  &  1163  &     2  &  1449  \\
     75  &   846  &     2  &  1113  \\
    100  &   707  &     2  &   901  \\
    150  &   538  &     1  &   647  \\
    200  &   443  &     0  &   505  \\
    300  &   336  &     0  &   320  \\
   1000  &   187  &     0  &    88  \\
   \hline
   Total &  4035  &  4753  &  4753  \\
   \hline
\end{tabular}                      
\end{center}
\end{minipage}
\end{table}

\begin{figure}
\centering
   \subfigure[\label{upsrat4i7ilow}]{
      \includegraphics[width=\columnwidth,clip]{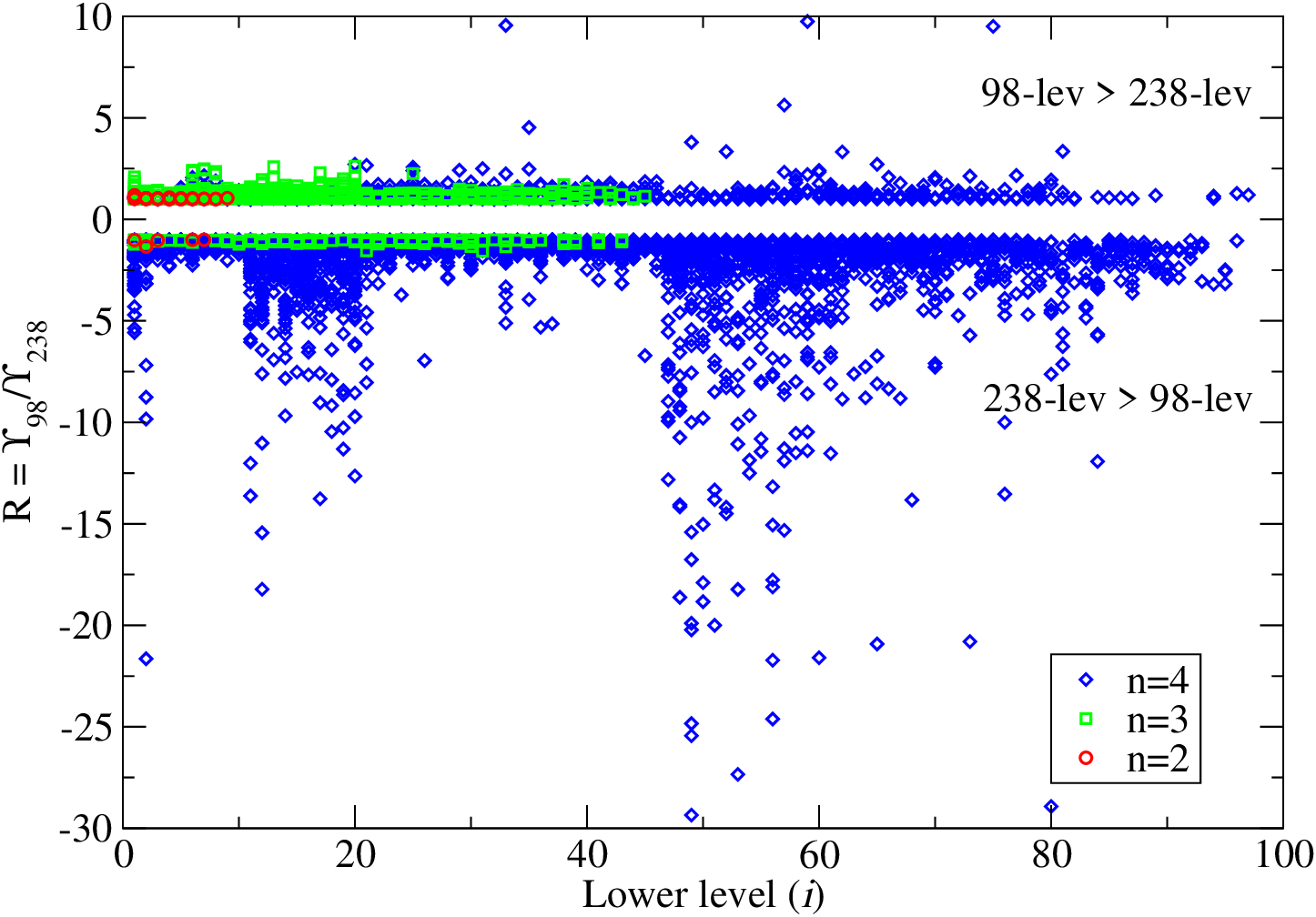}
   } \\
   \subfigure[\label{upsrat4i7iup}]{
      \includegraphics[width=\columnwidth,clip]{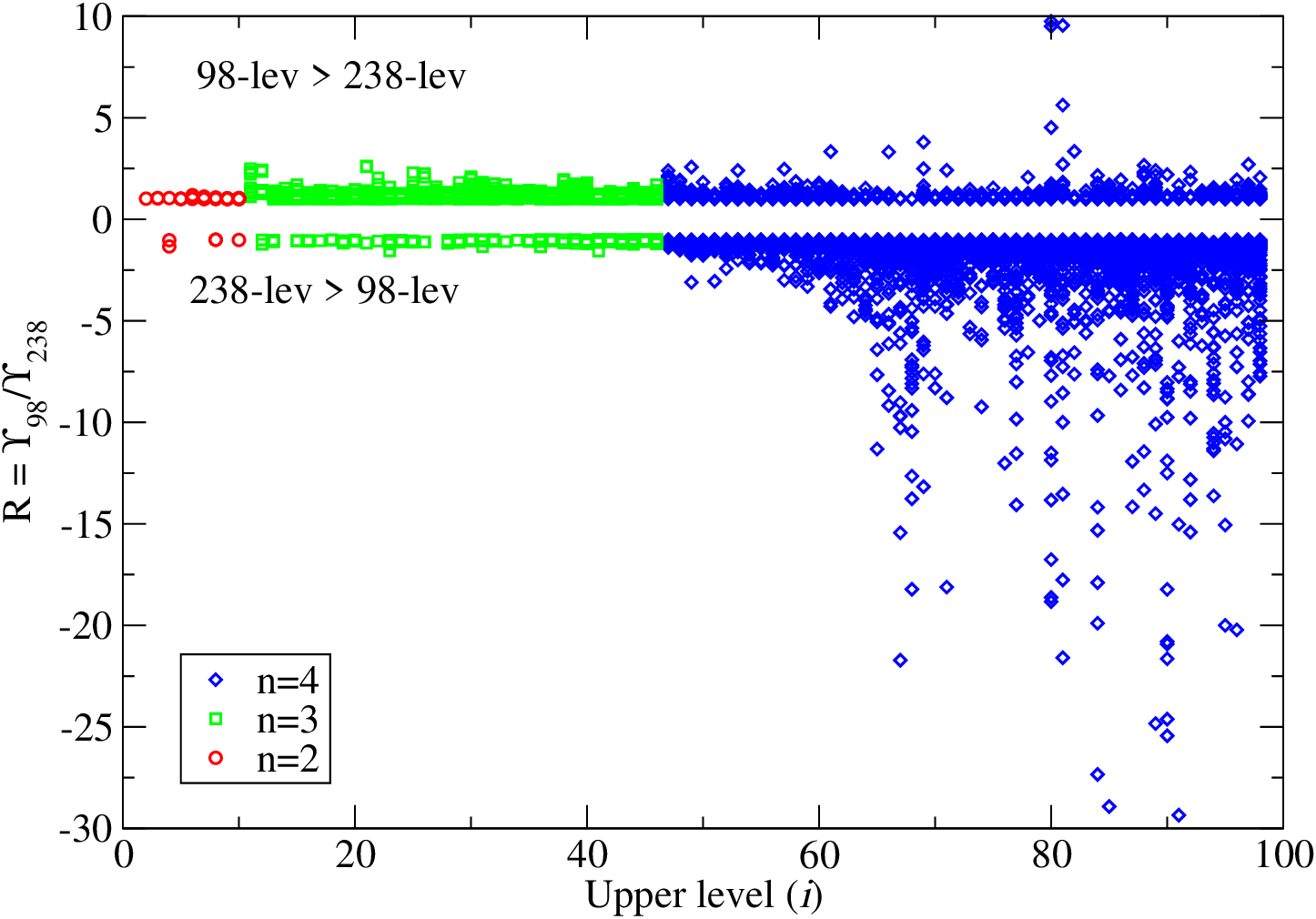}
   }
   \caption{The ratio of effective collision strengths from 98-
       vs 238-level CI/CC ICFT $R$-matrix calculations vs (a) lower level and 
      (b) upper level index,
      for all inelastic transitions amongst the 98 lowest common
      levels of Al~$^{9+}$ at
      $T_\mathrm{e} = 10^6 \Kelvin$.
      Positive values indicate 
      $\Upsilon_{98}>\Upsilon_{238}$ and negative values
      $\Upsilon_{238}>\Upsilon_{98}$.
      $\circ$: transitions with upper level  $n=2$; 
      $\square$: transitions with upper level  $n=3$; 
      $\diamond$: transitions with upper level  $n=4$.
      }
   \label{fig:upsrat4i7i_1e6K}
\end{figure}

The plots we have shown so far are useful in the respect that they allow us 
to make comparisons as a function of the strength of a transition --- larger
differences are acceptable for weaker transitions. 
On the other hand, because of the wide range of strengths, it is necessary for 
such plots to be logarithmic.
If we plot the ratio of results from two different calculations we can make a 
linear comparison. 
We do this in Fig.~\ref{fig:upsrat4i7i_1e6K} for the same comparison as we 
made in Fig.~\ref{fig:upsvups4i7i_1e6K}, and with respect to the lower level 
of the transition (\ref{upsrat4i7ilow}) or the upper one (\ref{upsrat4i7iup}).
Fig.~\ref{upsrat4i7ilow} shows that the transitions with the largest 
scatter are those from $2\mathrm{s}\,3l$  (lower level index 11--20) 
and $2\mathrm{s}\,4l$ (lower level index 47--60) up to $n=4$, and where we 
have highlighted by symbol / colour all upper levels with the same $n$-value.
It bears a very strong resemblance to the same comparison (figure 2) made by
\citet{aggarwal2015} to compare the 98-level CI/CC  DARC effective collision 
strengths with the 238-level CI/CC ICFT ones, except that they did not 
differentiate (highlight) the different $n$-values of the upper levels.
In contrast, \ref{upsrat4i7iup} clarifies that the differences in the
effective collision strengths become increasingly larger as the upper levels
excited move closer to the last one included in the 98-level CC calculation.
This in turn means that the effective collision strengths to the uppermost 
levels of the 238-level CC calculation of \citet{fernandez-menchero2014a} are 
increasingly unconverged with respect to the close-coupling expansion.
However, based on the present convergence study to $n=2$ and $n=3$ we
can expect that their results for transitions up to $n=4$ to be well 
converged, but those to $n=5$ less so since only a partial set of 
$n=6,7$ levels (to $l=2$) were included in their close-coupling expansion.

While all of the plots shown so far are visually appealing, especially 
Fig.~\ref{fig:upsrat4i7i_1e6K}, none of them give any indication of the
number of transitions whose quantities differ by any given amount --- we 
cannot tell the density of points close to the diagonal 
(Figs~\ref{fig:upsvups4i7i_inf}--\ref{fig:upsvups4l7l_1e6K}) or sitting
at unity (Fig.~\ref{fig:upsrat4i7i_1e6K}). 
We must be wary of such plots misleading us as to the level of agreement, as 
opposed to disagreement.
Only a table like Table \ref{tab:density} gives such an answer.

Thus, these comparisons demonstrate that the observation
by \citet{aggarwal2015}, that the 238-level CC effective collisions strengths 
of \citet{fernandez-menchero2014a} are systematically and increasingly larger 
with higher excitations than the  98-level CC results 
of \citet{aggarwal2014b} over a wide range of temperatures is correct, 
{\it but} it is due to the lack 
of convergence of the CC expansion of the 98-level CC results of 
\citet{aggarwal2014b}, particularly with respect to the $n=4$ levels.

\begin{figure}
\centering
   \includegraphics[width=\columnwidth,clip]{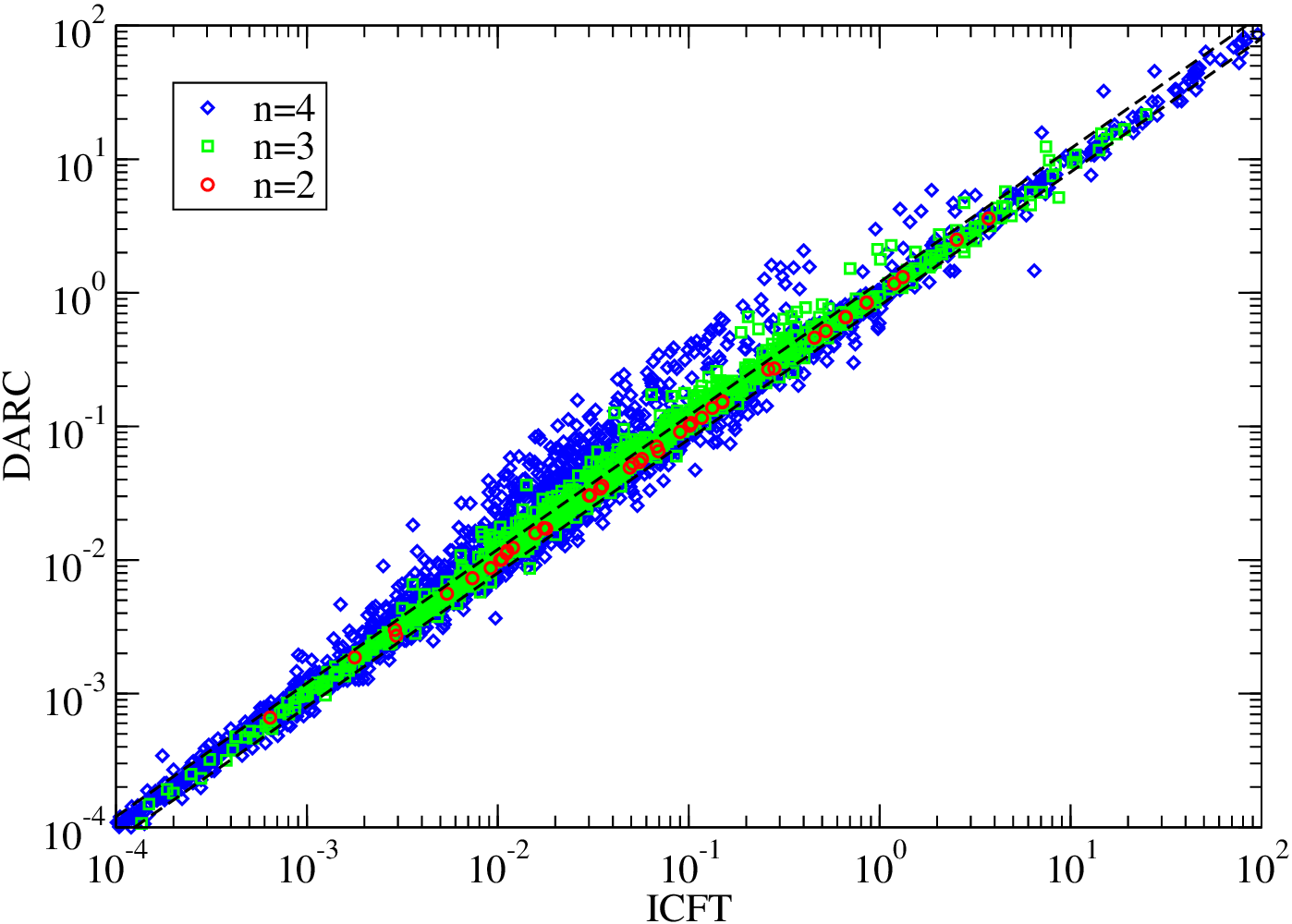}
   \caption{A comparison of effective collision strengths from 98-level CI/CC 
      ICFT versus Dirac $R$-matrix \citep{aggarwal2014b} calculations
      for all inelastic transitions amongst the 98 levels of Al~$^{9+}$
      at $T_\mathrm{e} = 10^6 \Kelvin$.
      $\circ$: transitions with upper level  $n=2$; 
      $\square$: transitions with upper level  $n=3$; 
      $\diamond$: transitions with upper level  $n=4$;
      dashed lines:  $20\%$ fractional difference.
      }
   \label{fig:upsvups4i4d_1e6K}
\end{figure}

Finally, there is in fact is good accord between comparable calculations,
viz. 98-level CI/CC, we make such a comparison of ICFT vs DARC in 
Fig.~\ref{fig:upsvups4i4d_1e6K}. The increasing  difference seen
as one progresses to higher levels is a reflection of the increasing
lack of convergence in the atomic structure. While both use the same
CI expansion, there is no reason for both to give the same unconverged result.
It is interesting to note that the weakest transitions, mostly forbidden ones,
show better agreement than some of the stronger allowed ones.
\citet{aggarwal2015} have already made a detailed comparison of 
98-level CI/CC DARC and 238-CI/CC ICFT effective collisions strengths for
transitions from the ground state. They highlighted several transitions to
$n=4$ which were discrepant, particularly at high temperatures. We consider
them in detail in Sec.~\ref{subsec:hightemp}

\subsection{Low temperature}
\label{subsec:lowtemp}

The effective collision strengths that we have presented so far have been 
relevant to the temperature of peak abundance (of Al~$^{9+}$) in an
electron collisional plasma, such as found in the solar atmosphere and 
magnetic fusion devices.
In photoionized plasmas the same charge state exists at much lower 
temperatures.
The role of resonances becomes more important at low temperatures both with 
respect to their magnitude, resolution and, in particular, their position.
\citet{fernandez-menchero2014a} carried-out an exhaustive analysis of 
the convergence of the effective collision strengths with respect to the 
energy step used to map-out the resonances.
They calculated the effective collision strengths $\Upsilon$ by convolution of 
the ordinary collision strengths $\Omega$ with a Maxwellian electron energy 
distribution.
Then they reduced the energy step by a factor one half and 
re-calculated $\Upsilon$.
After repeatedly reducing  the energy step down to a factor one eighth from the 
original one, the worst case transition from the ground state, the $1-78$, 
only changed by 10\% relative to the previous value of $\Upsilon$, at the 
lowest temperature calculated of $2 \times 10^4 \Kelvin$, and 
by $1\%$ at $10^6 \Kelvin$.

We note that \citet{aggarwal2015} incorrectly reported the energy step used in 
the resonance region by \citet{fernandez-menchero2014a}. 
As they stated, \citet{fernandez-menchero2014a} used an energy step in the 
resonance region that scales as $\sim z$ in the ion charge, not $z^2$, along 
the sequence.
In particular the step length used was 
$6.94 \times 10^{-6} z^2 \Ry$ for Al~$^{9+}$,
$4.89 \times 10^{-6} z^2 \Ry$ for $\mathrm{Cl}^{13+}$,
$4.09 \times 10^{-6} z^2 \Ry$ for $\mathrm{K}^{15+}$,
$3.30 \times 10^{-6} z^2 \Ry$ for $\mathrm{Ti}^{18+}$,
and $2.02 \times 10^{-6} z^2 \Ry$ for $\mathrm{Ge}^{28+}$.
These steps are comparable to the ones used 
by \citet{aggarwal2012a,aggarwal2014a,aggarwal2014b}, being slightly finer
for Al~$^{9+}$, and slightly coarser for $\mathrm{Ge}^{28+}$.
\citet{fernandez-menchero2014a} used this fine mesh only over $2J=1-23$, 
which corresponds to their exchange calculation.  
They used their coarse energy mesh across the resonance region as well 
for $2J=25-89$. 
Such $J$-values can only give rise to high-$n$ resonances, which by definition 
are narrow. 
A simple calculation with {\sc autostructure} reveals that the strongest 
resonances have widths $<10^{-5}$~Ryd.
Thus, the results provided by \citet{fernandez-menchero2014a} are converged 
with respect to the collision energy step and all significant resonances are 
well resolved.
Differences in low temperature effective collision strengths between
\citet{aggarwal2014b} and \citet{fernandez-menchero2014a}
can not be ascribed to the resolution of the resonances.

The largest source of error in the low temperature effective collision 
strengths arises from the inaccuracy in the positioning of resonances which 
lie just above threshold when the temperature (in energy units) starts to 
become comparable in magnitude with the uncertainty in position of these 
resonances. 
To a first approximation, this uncertainty in position is given by the 
difference between the calculated and observed values for the energy level to 
which the resonance is attached. 
In general, the specific level is not know, without detailed resonance 
analysis. 
%But, an estimate can be made based-on on typical errors for the likely levels 
%concerned.
%Energy level errors of 0.01 Ryd start to become important below a few 
%times $10^4 \Kelvin$.
Energy level accuracy can be improved theoretically via the use of 
pseudostates or purely `experimentally' through the use of observed energies 
or by a combination of theory and observation using term energy corrections. 
Each has its limitations: the use of pseudostates can lead to pseudoresonances 
at higher energies while not all levels may be known observationally. 
\citet{storey2014b} discuss the various considerations that need to be made 
in order to  calculate accurate data at very low temperatures for planetary nebulae,
for  example.

\begin{figure}
\centering
   \includegraphics[width=\columnwidth,clip]{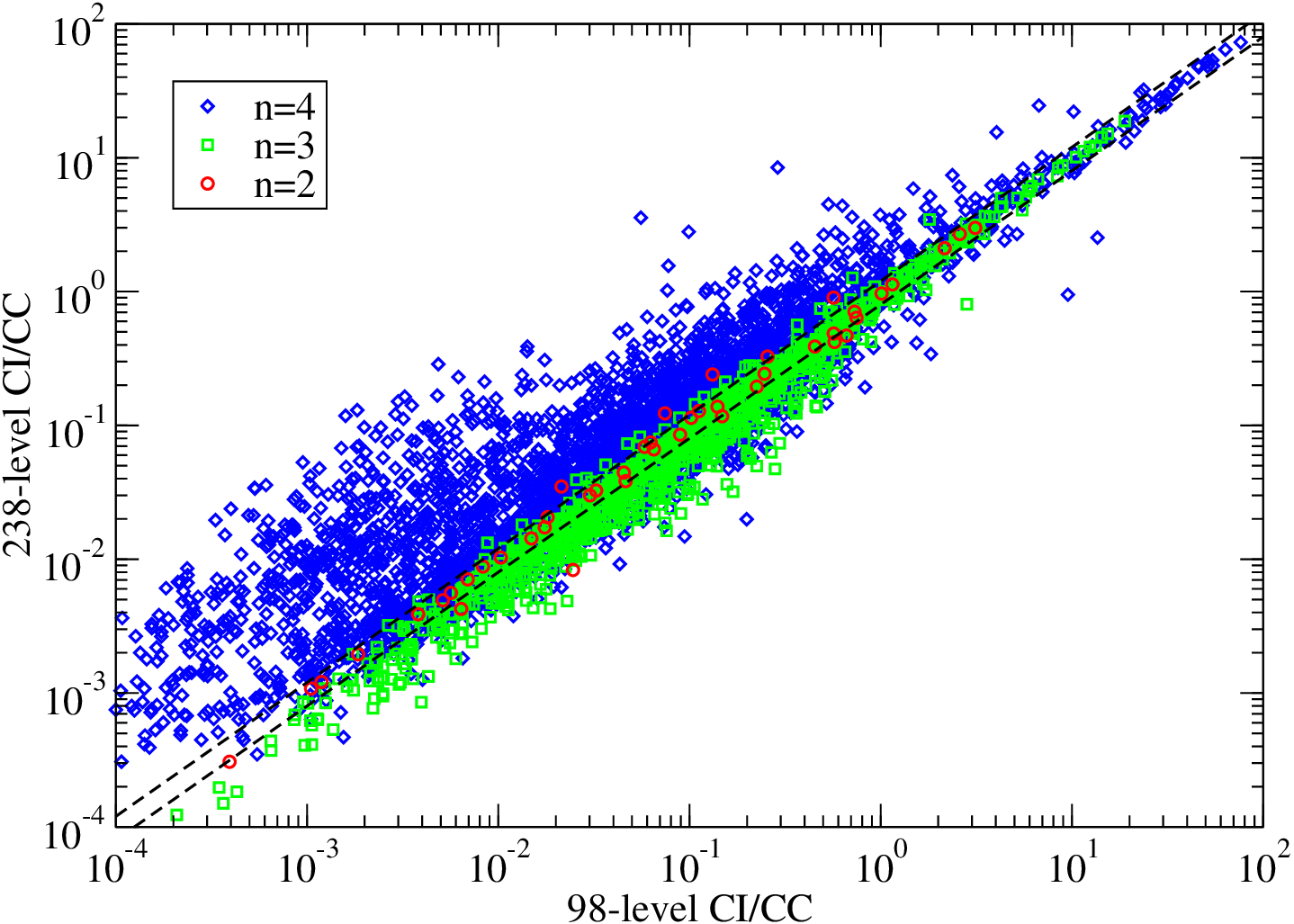}
   \caption{A comparison of effective collision strengths from 98-
      vs 238-level CI/CC ICFT $R$-matrix calculations for all inelastic 
      transitions amongst the 98 lowest common levels 
      of Al~$^{9+}$ at $T_\mathrm{e} = 2 \times 10^4 \Kelvin$.
      $\circ$: transitions with upper level  $n=2$; 
      $\square$: transitions with upper level  $n=3$; 
      $\diamond$: transitions with upper level  $n=4$;
      dashed lines:  $20\%$ fractional difference.
      }
   \label{fig:upsvups4i7i_2e4K}
\end{figure}

Fig.~\ref{fig:upsvups4i7i_2e4K} shows a comparison of 98- and 238-level CC 
ICFT effective collision strengths for transitions between the lowest common 
98 levels of Al~$^{9+}$ at a photoionized plasma temperature
of $2 \times 10^4 \Kelvin$ \citep{kallman2001}.
The same energy grid was used in both cases in the resonance region.
Differences are much larger than the ones seen at the temperature of peak 
abundance in an electron collisional plasma, even for transitions between the 
low-lying levels ($n=2$).
The increased differences in the effective collision strengths is likely due 
to the position of the resonances, especially where the 238-level CC effective 
collision strengths are smaller than the 98-level ones.
In addition, the systematic enhancement of the effective collision strengths 
to $n=4$ levels due to resonances attached to $n>4$ is increased due to the 
greater relative contribution from resonances;
i.e. the lack of convergence of the close-coupling expansion for these levels
in the 98-level calculation becomes even more significant.
However, at low temperatures, most modelling applications involve transitions
from the ground state, and perhaps a metastable. 
The factor $\exp(-\Delta E/kT)$ arising in the excitation rate coefficient 
means that only the lowest few excited levels are of interest, i.e. 
transitions within $n=2$.  

The inaccuracy in the position of the resonances makes effective collision strengths
from both \cite{fernandez-menchero2014a} and 
\citet{aggarwal2012a,aggarwal2014a,aggarwal2014b}
increasingly unreliable at low temperatures. 
For example, if we shift all resonances down in energy by $0.002$~Ryd 
(comparable with the accuracy of some energy levels) then the effective 
collision strength for the $1-4: \mathrm{2s^2\,^1S_0 - 2s2p\,^3P_2^o}$ 
transition changes by a factor of 2 at $2\times 10^4\Kelvin$, although this is 
reduced to a $20\%$ effect by $2\times 10^5\Kelvin$.
Nevertheless, such results can be used for estimation purposes.

\subsection{High temperature}
\label{subsec:hightemp}

We calculate effective collision strengths over a wide range of
temperatures, as defined by the OPEN-ADAS {\it adf04} file format
viz. $2\times 10^2 - 2\times 10^6 \; (z+1)^2$~K, to cover all possible applications
to electron-collisional plasmas. 
Higher temperature effective collision strengths require ordinary collision 
strengths to higher energies, which in turn require the contribution from 
higher partial waves, and this needs to be handled efficiently and accurately
by $R$-matrix calculations.

\citet{aggarwal2015} highlighted several transitions ($1-64$, $1-70$ and $1-80$)
for which they observed large differences between the 238-level CI/CC ICFT 
results of  \citet{fernandez-menchero2014a} and the 98-level CI/CC DARC ones
 of \citet{aggarwal2014b}, particularly at high temperatures. They suggested that the 
use of the \citet{burgess1992} formulae at high energy was perhaps 
a major source of error, i.e. that the ICFT calculations did not go high enough in 
energy for the collision strengths to have reached their asymptotic form.
\citet{aggarwal2015} also commented-on the neglect of electron exchange
by \citet{fernandez-menchero2014a} at high-$J$.

The `top-up' procedure used for angular momentum is described 
in \citet{fernandez-menchero2014a}.
In addition, \citet{fernandez-menchero2014a} included electron exchange for 
angular momenta up to $2J=23$, and then used a non-exchange calculation for 
the rest of the angular momenta calculated: $2J=25-89$.
(\citet{aggarwal2014b} included exchange for all of the angular momenta 
calculated, up to $2J=91$.)
The method used by \citet{fernandez-menchero2014a}
 is not a source of significant inaccuracy. 
By $2J=25$ the smallest exchange multipole is larger than $10$.  
Neglect of higher exchange multipoles causes a small underestimate at the 
highest temperatures for a few very weak highly forbidden transitions, 
i.e. ones that not only have no target mixing with  allowed transitions
(i.e. zero limit points) but also are not strongly enhanced by coupling. 
By extending the inclusion of exchange to $2J=51$ we find no transition 
differing by more than $5\%$ up to $2\times 10^7 \Kelvin$,
rising to $10\%$ at $2\times 10^8 \Kelvin$, for Al~$^{9+}$.

With regards to energy, \citet{fernandez-menchero2014a} extended the outer 
region $R$-matrix calculation up to three times the ionization potential, 
$88\,\mathrm{Ry}$ in the case of Al~$^{9+}$.
They then carried-out a linear interpolation of the reduced collision 
strength, $y$, as a function of the reduced scattering energy, $x \in [0,1]$, 
for dipole and Born allowed transitions (while forbidden transitions were 
extrapolated) as follows.

For a given excitation, let $E$ denote the final scattered energy 
(with $\Delta E$  the excitation  energy still) and define
\begin{equation}
   \varepsilon\ =\ \frac{E}{\Delta E}\,.
\end{equation}
Then, at threshold ($E=0$) $\varepsilon=0$.
Following \citet{burgess1992}, we divide all transitions into one of three 
cases to represent $(x,y)$, based-on their infinite energy values $y_\infty$, 
or lack thereof, as described in sec.~\ref{sec:structure}.

\begin{itemize}
   \item Dipole transitions
\begin{eqnarray}
   x & = & 1\ -\ \frac{\ln C}{\ln \left( \varepsilon\,+\,C \right)}
           \nonumber \\
   y & = & \frac{\Omega}{\ln \left( \varepsilon\,+\,\re \right)}
\label{eq:burgdipole}
\end{eqnarray}
   \item Born transitions
\begin{eqnarray}
   x & = & \frac{\varepsilon}{\varepsilon\,+\,C}
           \nonumber \\
   y & = & \Omega
\label{eq:burgborn}
\end{eqnarray}
   \item Forbidden transitions
\begin{eqnarray}
   x & = & \frac{\varepsilon}{\varepsilon\,+\,C}
           \nonumber \\
   y & = & \left( \varepsilon\,+\,1 \right)^{\alpha}\,\Omega
\label{eq:burgforbidden}
\end{eqnarray}
\end{itemize}
where $C$ is an arbitrary visual scaling parameter.
In the last case, formally, $\alpha=2$ in the infinite energy limit.
At high but finite energies, a more accurate approach is to determine
$\alpha$ from two reasonably well-separated high-energy points so as to take
account of enhancement or depletion of these weak high-energy collision 
strengths by continuum coupling. 
This is carried-out automatically, but restricted to the range $\alpha=1-3$, 
i.e. within reasonable physical bounds.

\begin{figure}
\centering
   \subfigure[\label{fig:al9omgburg_1-52} 
      $1-52: \mathrm{2s^2\,^1S_0 - 2s4p\,^1P_1^o}$ ]{
      \includegraphics[width=0.45\columnwidth,clip]{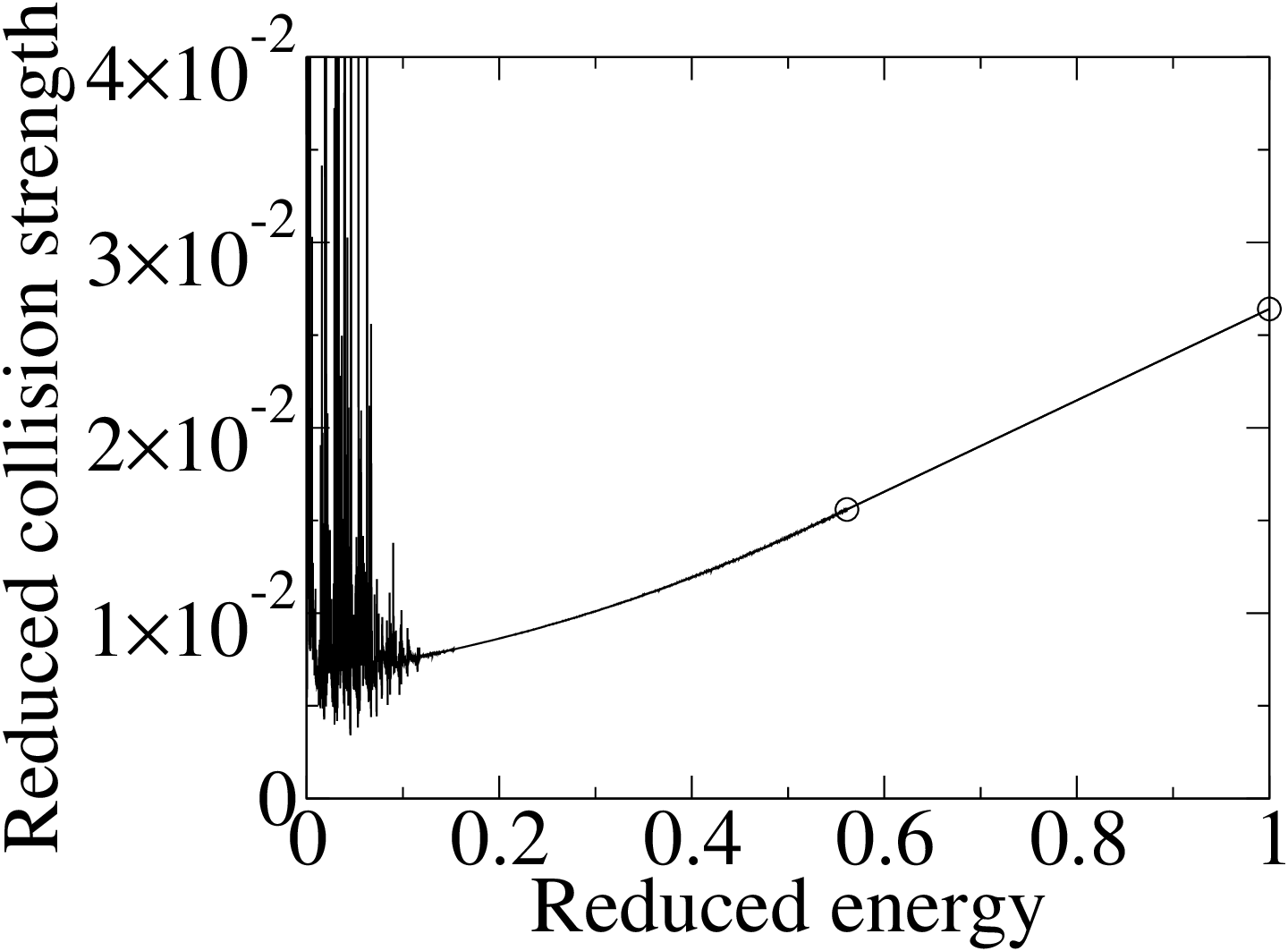}
   } 
   \subfigure[\label{fig:al9omgburg_1-64} 
      $1-64: \mathrm{2s^2\,^1S_0 - 2p4s\,^1P_1^o}$;]{
      \includegraphics[width=0.45\columnwidth,clip]{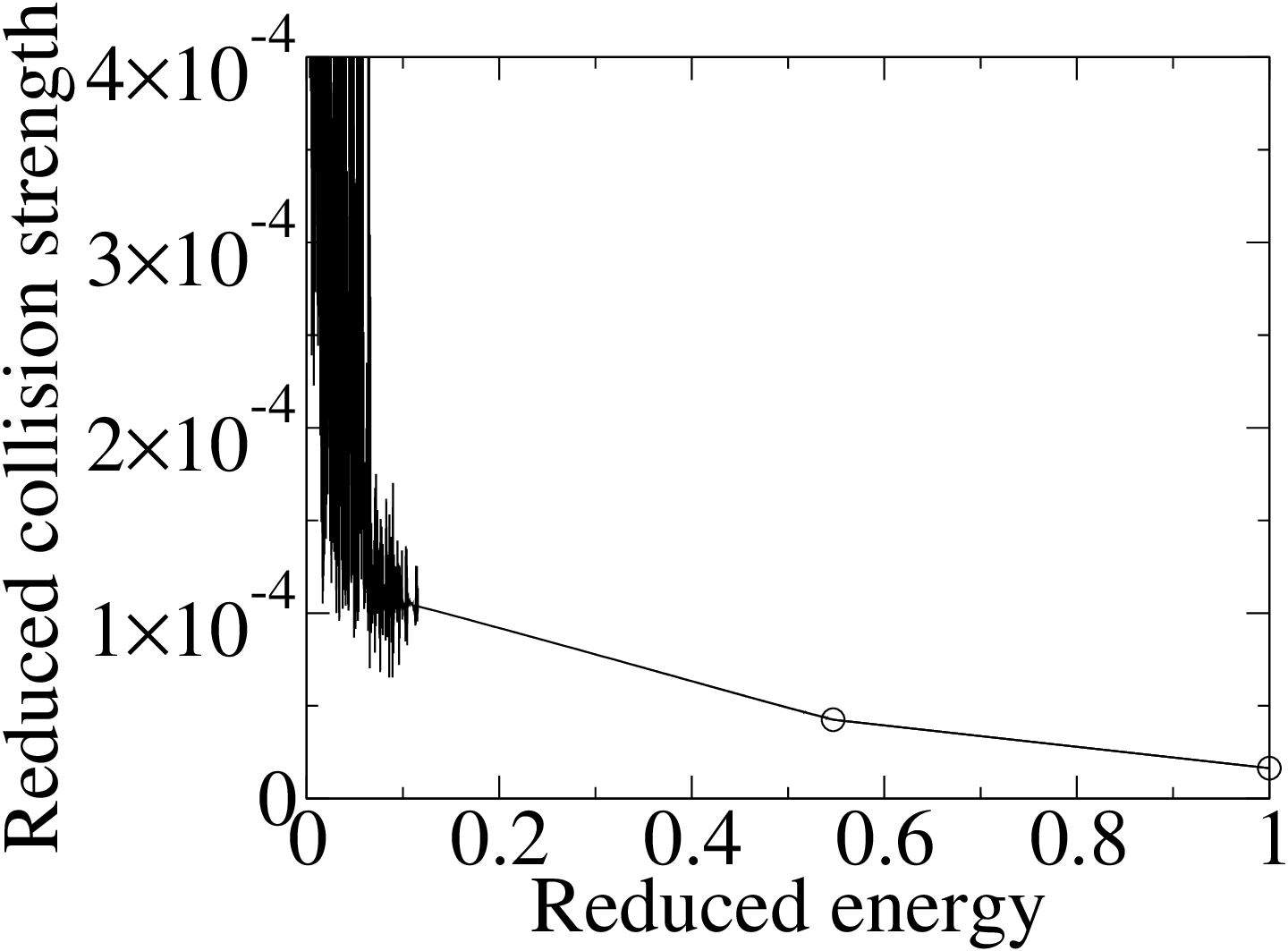}
   } \\
   \subfigure[\label{fig:al9omgburg_1-70} 
      $1-70: \mathrm{2s^2\,^1S_0 - 2p4p\,^3P_0}$]{
      \includegraphics[width=0.45\columnwidth,clip]{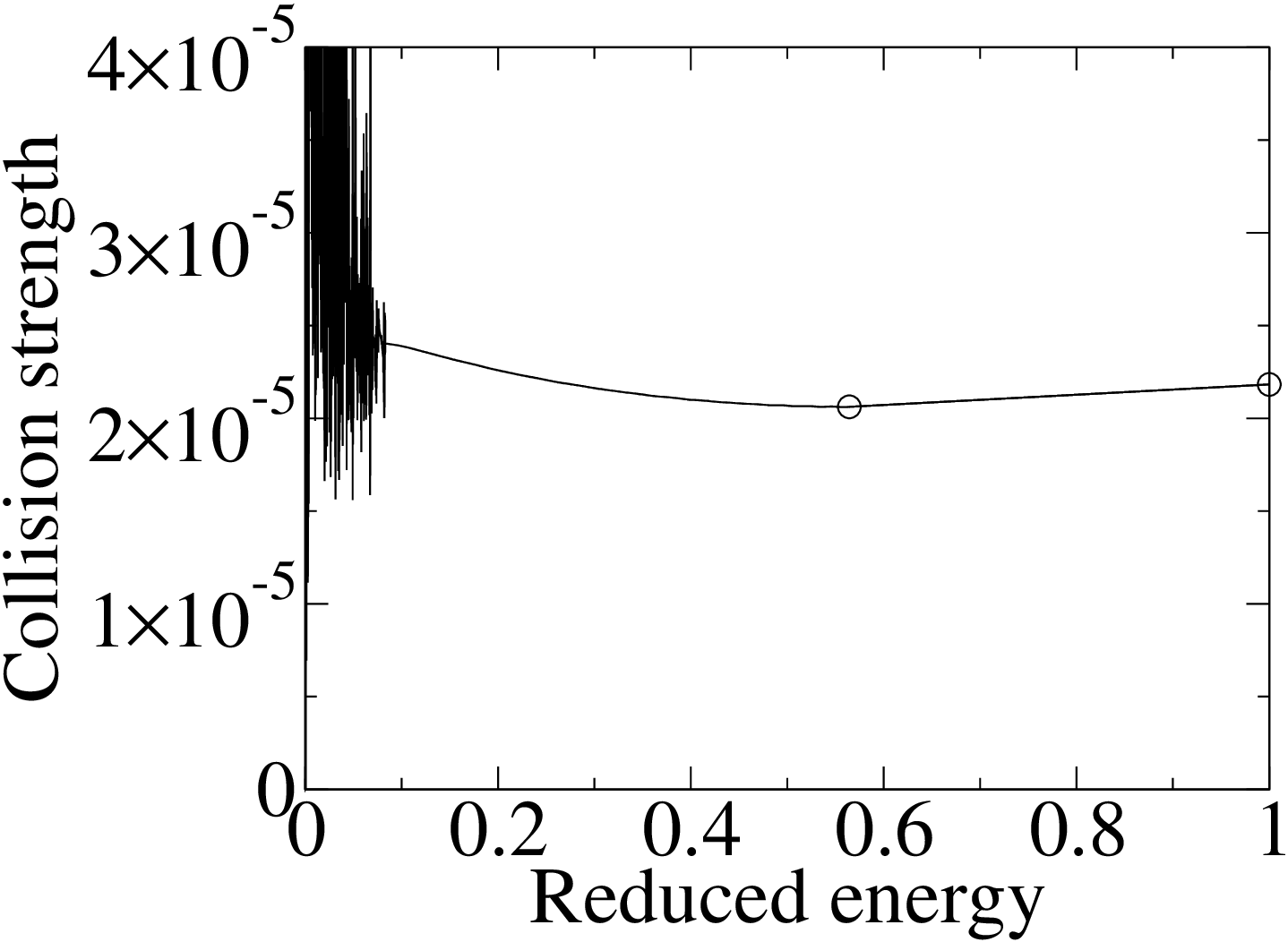}
   } 
   \subfigure[\label{fig:al9omgburg_1-80} 
      $1-80: \mathrm{2s^2\,^1S_0 - 2p4f\,^1F_3}$]{
      \includegraphics[width=0.45\columnwidth,clip]{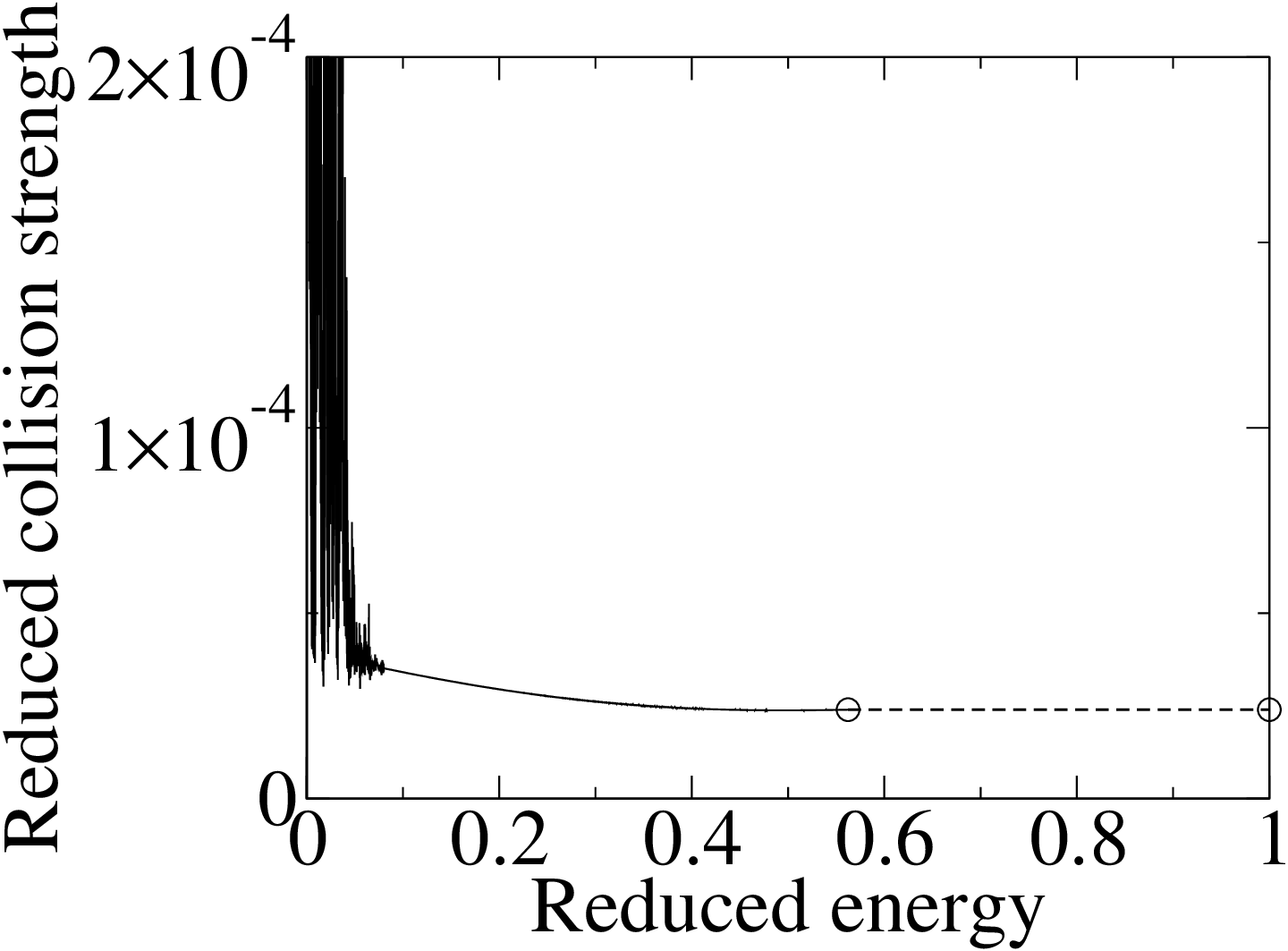}
   } \\
   \caption{Reduced collision strengths versus energy from 238-level CI/CC 
      ICFT $R$-matrix calculations for selected transitions 
      of Al~$^{9+}$.
%(a) $1-52: \mathrm{2s^2\,^1S_0 - 2s4p\,^1P_1^o}$;
%(b) $1-64: \mathrm{2s^2\,^1S_0 - 2p4s\,^1P_1^o}$;
%(c) $1-70: \mathrm{2s^2\,^1S_0 - 2p4p\,^3P_0}$;
%(d) $1-80: \mathrm{2s^2\,^1S_0 - 2p4f\,^1F_3}$.
      $\circ$: denotes the last finite calculated and infinite energy points;
      dashed line: indicates extrapolation for a forbidden transition. ($C=2.$)}
   \label{fig:al9omgburg}
\end{figure}

Fig.~\ref{fig:al9omgburg} shows the reduced collision strength in a 
Burgess--Tully $(x,y)$ plot from the 238-level CI/CC ICFT calculation for 
Al~$^{9+}$ by \citet{fernandez-menchero2014a}.
Fig.~\ref{fig:al9omgburg_1-52} is for a strong dipole transition 
$1-52:\mathrm{2s^2\,^1S_0 - 2s4p\,^1P_1^o}$,
Fig.~\ref{fig:al9omgburg_1-64} the dipole transition
$1-64: \mathrm{2s^2\,^1S_0 - 2p4s\,^1P_1^o}$ which takes 
place through configuration mixing,
Fig.~\ref{fig:al9omgburg_1-70} the Born-allowed transition
$1-70: \mathrm{2s^2\,^1S_0 - 2p4p\,^3P_0}$ which
also takes place only through configuration mixing, and 
Fig.~\ref{fig:al9omgburg_1-80} is for the forbidden transition 
$1-80: \mathrm{2s^2\,^1S_0 - 2p4f\,^1F_3}$ which is a very weak two-electron 
jump. 
An automatically determined value of $\alpha=1$ for this transition was used 
to extrapolate the reduced collision strength as a function of reduced energy 
--- see equation~\ref{eq:burgforbidden}.
Fig.~\ref{fig:al9omgburg} shows that all of the transitions have reached the 
assumed asymptotic form.
What then is the source of the differences in high temperature effective 
collision strengths noted by \citet{aggarwal2015}?
The answer lies in the atomic structure. 

In Table \ref{tab:upshighT} we compare effective collision strengths 
from the 98- and 238-level CI/CC ICFT calculations with the 98-level 
CI/CC DARC ones  \citep{aggarwal2014b}.   
We note first that results for the strong dipole
transition $1-21$ are independent of atomic structure (98- vs 238-level CI 
$y_\infty=4S/3$) and close-coupling expansion (98- vs 238-level CC $\Upsilon$). 
However, if we consider the weak dipole transition $1-64$ we see that the limit
value ($y_\infty$) is a factor 6.2 larger for the 98- vs 238-level CI (Breit--Pauli) 
case and this leads to a factor of 2.08 in the corresponding ICFT effective
collision strengths at $2 \times 10^{7}\Kelvin$. Indeed, the difference
in effective collision strengths would likely be larger were it not for the
fact that the 238-level CI/CC is (much) larger at (much) lower temperatures
due to additional resonances and coupling.
The DARC structure limit point reported by \citet{aggarwal2015} is similar 
to the 98-level CI Breit--Pauli one. Correspondingly, the 98-level CI/CC DARC and ICFT 
effective collision strengths agree to within $\sim 20\%$ over the entire 
temperature range shown in Table \ref{tab:upshighT}.

For the case of the (weak) Born-allowed transition $1-70$ we see a similar
trend in the comparisons, viz. differences in structure ($y_\infty$) leading to 
corresponding differences in high temperature effective collision strengths,
strong resonance enhancement at lower temperature for the 238-  vs 98-level
CI/CC results and agreement to within 30\% between the DARC and ICFT 98-level CI/CC
results. (\citet{aggarwal2015} do not report Born limits for this transition,
but clearly the sensitivity to atomic structure we see reflected in the two 
Breit--Pauli Born limits accounts for the remaining difference.) 
Finally, for the weak forbidden $1-80$ transition we note a very similar  
set of comparisons as for the $1-70$ transition, indeed, the DARC and ICFT 
98-level CI/CC results agree more closely (20\%).
The 238-level CI/CC ICFT result is increasingly enhanced by resonances over the 
98-level CI/CC one at low temperatures, by a factor 15 at $2\times 10^4$~K.
We note that including higher-$J$ exchange multipoles does not change the
effective collision strength to 3 s.f. even at the highest temperature considered.
There are several equally forbidden transitions ($1-61, 68, 69$, not shown) for
which the pattern of dis/agreement is very similar to that for the $1-80$, in
all cases.

In summary, the results of the 98-level CI/CC DARC 
calculation  of \citet{aggarwal2014b} are in much closer agreement (indeed, 
no significant  differences) with the present 98-level CI/CC ICFT results 
than the 238-level CI/CC ones across  a wide range of  temperatures 
for all of the transitions highlighted by \citet{aggarwal2015}.
However, the results of the calculations obtained using the 238-level CI 
target have a better converged atomic structure and, correspondingly, 
give more accurate effective collision strengths, especially at high 
temperatures, while the much better convergence of the 238-level 
close-coupling expansion provides more accurate results across a wide 
range of temperatures.

\begin{table*}
\begin{minipage}{\textwidth}
\caption{\label{tab:upshighT} Comparison of effective collision strengths, 
   $\Upsilon$, at selected finite temperatures, and the reduced quantity 
   $y_\infty$ at infinite temperature.
   See Table \ref{tab:energies} for the transition indices $i, j$.}
\begin{center}
\begin{tabular}{r@{--}l ccc c cccc c cccc}
   \hline
   \multicolumn{2}{l}{}      & %Transition
   \multicolumn{3}{c}{DARC}       & &
   \multicolumn{9}{c}{ICFT} \vspace{0.1cm}  \\ 
   \multicolumn{2}{c}{} \vspace{-0.4cm} \\ 
   \cline{7-15}
   \multicolumn{2}{c}{} \vspace{-0.2cm} \\ 
   $i$  & $j$ &
   \multicolumn{3}{c}{98-level CI/CC}  & &
   \multicolumn{4}{c}{98-level CI/CC}  & &
   \multicolumn{4}{c}{238-level CI/CC} \\ 
   \cline{3-5} \cline{7-10} \cline{12-15}
   \multicolumn{2}{c}{} \vspace{-0.2cm} \\ 
   \multicolumn{2}{l}{$T$(K)} &
   $2 \times 10^{4}$ & $10^{6}$ & $2 \times 10^{7}$ & &
   $2 \times 10^{4}$ & $10^{6}$ & $2 \times 10^{7}$ & $\infty$ & &
   $2 \times 10^{4}$ & $10^{6}$ & $2 \times 10^{7}$ & $\infty$  \\
   \hline
    1 & 52 & $1.09\,(-2)$ & $1.01\,(-2)$ & $3.01\,(-2)$ & &
             $1.14\,(-2)$ & $1.02\,(-2)$ & $3.20\,(-2)$ & $2.40\,(-2)$ & &
             $1.53\,(-2)$ & $9.68\,(-3)$ & $3.22\,(-2)$ & $2.64\,(-2)$ \\
    1 & 64 & $5.49\,(-4)$ & $1.78\,(-4)$ & $1.70\,(-4)$ & &
             $5.80\,(-4)$ & $1.80\,(-4)$ & $1.96\,(-4)$ & $1.01\,(-4)$ & &
             $3.12\,(-3)$ & $2.94\,(-4)$ & $9.40\,(-5)$ & $1.63\,(-5)$ \\
    1 & 70 & $2.68\,(-5)$ & $1.26\,(-5)$ & $5.29\,(-6)$ & &
             $3.48\,(-5)$ & $1.11\,(-5)$ & $4.24\,(-6)$ & $2.68\,(-6)$ & &
             $6.99\,(-5)$ & $3.10\,(-5)$ & $2.18\,(-5)$ & $2.18\,(-5)$ \\
    1 & 80 & $4.52\,(-5)$ & $2.51\,(-5)$ & $8.45\,(-6)$ & &
             $4.63\,(-5)$ & $2.47\,(-5)$ & $7.05\,(-6)$ & $-$          & &
             $7.04\,(-4)$ & $8.66\,(-5)$ & $1.09\,(-5)$ & $-$          \\
   \hline
\end{tabular}
\end{center}
   Notes. 
   DARC: \citet{aggarwal2014b};
   98-level ICFT: present work;
   238-level ICFT: \citet{fernandez-menchero2014a}. %\\
   $A\,(B)$ denotes $A \times 10^{B}$.
\end{minipage}
\end{table*}

\section{Conclusions}
\label{sec:conclusions}

Reliable and accurate electron-impact excitation data are key to the 
successful spectroscopic diagnostic modelling of non-LTE plasmas.
We have compared and contrasted differences in such data for the benchmark 
Be-like Al~$^{9+}$ ion which we have calculated using the $R$-matrix method.
Such differences arise through:
1) differing approximations of relativistic effects, 
2) uncertainties in atomic structure and
3) errors due to the lack of convergence of the close-coupling expansion.
Error 3) is quantifiable and can be reduced systematically and reliably ---
we  illustrated this by comparing new 98-level and previous 238-level CC
ICFT $R$-matrix calculations.
We find that effective collision strengths to $n=4$ levels are significantly
enhanced over a wide range of temperatures by coupling to $n>4$ levels.
Uncertainty 2) is quantifiable but is more difficult to reduce and
constrain as an error --- we compared 98-level and 238-level configuration 
interaction expansion calculations of line strengths and infinite energy 
plane-wave Born collision strengths to illustrate this point.
Again, transitions to $n=4$ levels are most susceptible to lack of
convergence but now of the configuration interaction expansion.
Differences 1) between ICFT and Breit-Pauli $R$-matrix treatments of
relativistic effects are small, and negligible relative to 2) and 3),
as is to be expected for an element which lies below Zn. 
We illustrated this by a comparison of new 98-level CI/CC ICFT and 
Breit--Pauli $R$-matrix effective collision strengths which use the exact same 
atomic structure. 
We also find good accord between our 98-level CI/CC results and previous ones
from a 98-level CI/CC Dirac--Coulomb $R$-matrix calculation, particularly
for transitions from the ground-level. 

However, based-upon the study of effects 1), 2) and 3), we conclude that the original
238-level CI/CC ICFT $R$-matrix results  are the most complete to-date with  
respect to convergence of both the configuration interaction and close-coupling
expansions and a reliable treatment of relativistic effects.
Or to put it more simply,  the earlier 238-level CI/CC ICFT work 
\citep{fernandez-menchero2014a} has a superior target to the 98-level CI/CC DARC 
one \citep{aggarwal2014b} and provides more accurate atomic data.

Thus, we find to be false the recent conjecture by \citet{aggarwal2015}
that the ICFT approach may  not be completely robust. Their conjecture was based upon
a comparison of  98-level CI/CC Dirac $R$-matrix effective collision strengths 
\citep{aggarwal2014b} with those from 238-level CI/CC ICFT $R$-matrix calculations
\citep{fernandez-menchero2014a}. Rather, \citet{aggarwal2015}
have failed to appreciate the size of the effect which the lack of convergence,
in both the close-coupling and configuration interaction expansions,
has on transitions to the higher-lying states ($n=4$ in the case of a 
98-level CI/CC expansion for Al~$^{9+}$). This can only be quantified
by extending the expansions. 

There is nothing special about Al~$^{9+}$ with regards to the
convergence of the close-coupling expansion. 
Except perhaps at the lowest charge states, as they discussed, the effective 
collision  strengths for all ions in the $\mathrm{Be}$-like sequence, 
from $\mathrm{B}^{+}$ to $\mathrm{Kr}^{32+}$ calculated 
by \cite{fernandez-menchero2014a}\footnote{ 
   Fig.~5 of the paper of \citet{fernandez-menchero2014a} compared their
   effective collision strengths for the transition 
   $\mathrm{2s^2\,^1S_0 - 2s2p\,^3P_1}$ of $\mathrm{P}^{11+}$ with the 
   corresponding interpolated results from~\citet{keenan1988b}.
   The fitting coefficients of \citet{keenan1988b} were taken in numerical 
   form from CHIANTI v7.1 rather than being transcribed from the original 
   paper. 
   However, one of the coefficients was missing a sign, as pointed out 
   by \citet{aggarwal2015}.
   This will be corrected in the next release of CHIANTI.
   The calculated results of \citet{fernandez-menchero2014a} for this 
   transition are now only $50\%$ larger than the interpolated 
   ones \citep{keenan1988b} at the temperature of peak abundance 
   ($1.8\times 10^6$~K), for example.
}
using the 238~level CI/CC expansion are the most complete and reliable 
to-date. 
The use of a 98~level CI/CC expansion for Al~$^{9+}$
\citep{aggarwal2014b}, $\mathrm{Cl}^{13+}$, $\mathrm{K}^{15+}$, 
$\mathrm{Ge}^{28+}$ \citep{aggarwal2014a} 
and $\mathrm{Ti}^{18+}$~\citep{aggarwal2012a} means that these effective 
collision strengths for transitions up to $n=4$ are increasingly an 
underestimate over a wide range of temperatures, including the temperature of 
peak abundance.

There is nothing special about $\mathrm{Be}$-like ions. 
The 136-level CC DARC calculations of \citet{aggarwal2014c} 
for $\mathrm{Fe}^{13+}$ are shown to be a systematic underestimate compared to the 
197-level CC ICFT calculations of \citet{liang2010b} (in addition, Liang et al. used
a much larger CI expansion 2985 vs 136 levels) --- 
see Del Zanna et al (MNRAS to be submitted) for another detailed analysis 
similar to the present paper's.
The convergence of the close-coupling expansion increasingly affects the 
accuracy of collision data to the highest-lying states in all $R$-matrix 
calculations. 
Likewise, the accuracy of the atomic structure becomes more uncertain for the 
most highly-excited states of the configuration interaction expansion. 

In general, care must be exercised when comparing collision data calculated 
using different atomic structures and / or close-coupling expansions lest one 
draws false conclusions. 
Finally, given Figs \ref{fig:upsvups4i7i_inf} and \ref{fig:upsvups4i7i_1e6K} 
and Table \ref{tab:density}, we suggest that it is fanciful to assign a 
single accuracy rating of $20\%$, say, to an entire collision data set.

\section*{Acknowledgments}
The present work was funded by STFC (UK) through the 
 University of Strathclyde UK APAP network grant
ST/J000892/1 and the 
University of Cambridge DAMTP astrophysics grant. 

\def\baselinestretch{1.0}

\bibliographystyle{mn2e}
\bibliography{references}

\appendix

\section{C$^{2+}$}

Just recently, another paper by 
\citet{aggarwal2015b} has appeared, this time on Be-like $\mathrm{C}^{2+}$,
making much the same claims that have just been refuted, quite generally, above.
\citet{aggarwal2015b} have extended their CI/CC expansion up to $n=5$ this time, 
but this 166-level expansion still falls short of the 238-level expansion up to 
$nl=7\mathrm{d}$ of \citet{fernandez-menchero2014a}. This may reduce the systematic
differences (understimates) of their results up to $n=4$ somewhat compared to those of 
\citet{fernandez-menchero2014a} but the low-charge state means that the errors
and uncertainties due to the difference between the two atomic structures will
be much larger than for $\mathrm{Al}^{9+}$. Arguably, a Breit--Pauli $R$-matrix with 
pseudo-states calculation is required for $\mathrm{C}^{2+}$ to give a definitive 
representation of the CI/CC expansion for all level-resolved transitions.
\label{lastpage}

\end{document}

%% file: energies.tex
\begin{table*}
\begin{minipage}{\textwidth}
   \caption{$\mathrm{Al}^{9+}$ target levels.
      Key: $i$: level index; Conf.: configuration; 
      Level: level IC designation (largest weight); 
      $E_{\mathrm{NIST}}$: observed energy from the NIST database (\citealt{martin1979});
      $E_{98}$: calculated energy with 98-level CI expansion;
      $E_{238}$: calculated energy with 238-level CI expansion (see text);
      \%: percentage difference between theoretical and NIST data.
      All energies are in $\mathrm{cm}^{-1}$. }
   \label{tab:energies} 
\begin{center}
\begin{small}
\begin{tabular}{rr@{\ }rrr@{\ (}r@{)\quad}r@{\ (}r@{)\qquad}rr@{\ }rrr@{\ (}r@{)\quad}r@{\ (}r@{)\quad}} 
   \hline
   $i$ & Conf.             & Level          & $E_{\mathrm{NIST}}$ & $E_{98}$       & \%     & $E_{238}$      & \%     &     $i$ & Conf.             & Level          & $E_{\mathrm{NIST}}$ & $E_{98}$       & \%     & $E_{238}$      & \%     \\ 
   \hline
   \hline
     1 & $\mathrm{2s^2  }$ & $\mathrm{^1S_{0  }    }$ & $        0.$ & $        0.$ & $   0$ & $        0.$ & $   0$ &      50 & $\mathrm{2s\,4p}$ & $\mathrm{^3P_{1  }^{o}}$ & $        - $ & $  2504724.$ & $  - $ & $  2500556.$ & $  - $ \\ 
     2 & $\mathrm{2s\,2p}$ & $\mathrm{^3P_{0  }^{o}}$ & $   155148.$ & $   155722.$ & $ 0.4$ & $   155539.$ & $ 0.3$ &      51 & $\mathrm{2s\,4p}$ & $\mathrm{^3P_{2  }^{o}}$ & $        - $ & $  2505125.$ & $  - $ & $  2500975.$ & $  - $ \\ 
     3 & $\mathrm{2s\,2p}$ & $\mathrm{^3P_{1  }^{o}}$ & $   156798.$ & $   157487.$ & $ 0.4$ & $   157404.$ & $ 0.4$ &      52 & $\mathrm{2s\,4p}$ & $\mathrm{^1P_{1  }^{o}}$ & $        - $ & $  2508073.$ & $  - $ & $  2503514.$ & $  - $ \\ 
     4 & $\mathrm{2s\,2p}$ & $\mathrm{^3P_{2  }^{o}}$ & $   160429.$ & $   161146.$ & $ 0.4$ & $   161278.$ & $ 0.5$ &      53 & $\mathrm{2s\,4d}$ & $\mathrm{^3D_{1  }    }$ & $        - $ & $  2520502.$ & $  - $ & $  2516248.$ & $  - $ \\ 
     5 & $\mathrm{2s\,2p}$ & $\mathrm{^1P_{1  }^{o}}$ & $   300490.$ & $   309273.$ & $ 2.9$ & $   307209.$ & $ 2.2$ &      54 & $\mathrm{2s\,4d}$ & $\mathrm{^3D_{2  }    }$ & $        - $ & $  2520573.$ & $  - $ & $  2516321.$ & $  - $ \\ 
     6 & $\mathrm{2p^2  }$ & $\mathrm{^3P_{0  }    }$ & $   404574.$ & $   408026.$ & $ 0.9$ & $   407826.$ & $ 0.8$ &      55 & $\mathrm{2s\,4d}$ & $\mathrm{^3D_{3  }    }$ & $        - $ & $  2520682.$ & $  - $ & $  2516434.$ & $  - $ \\ 
     7 & $\mathrm{2p^2  }$ & $\mathrm{^3P_{1  }    }$ & $   406517.$ & $   409969.$ & $ 0.8$ & $   409888.$ & $ 0.8$ &      56 & $\mathrm{2s\,4d}$ & $\mathrm{^1D_{2  }    }$ & $  2527560.$ & $  2530218.$ & $ 0.1$ & $  2525880.$ & $-0.1$ \\ 
     8 & $\mathrm{2p^2  }$ & $\mathrm{^3P_{2  }    }$ & $   409690.$ & $   413420.$ & $ 0.9$ & $   413526.$ & $ 0.9$ &      57 & $\mathrm{2s\,4f}$ & $\mathrm{^3F_{2  }^{o}}$ & $  2528570.$ & $  2530774.$ & $ 0.1$ & $  2526258.$ & $-0.1$ \\ 
     9 & $\mathrm{2p^2  }$ & $\mathrm{^1D_{2  }    }$ & $   449732.$ & $   458157.$ & $ 1.9$ & $   457831.$ & $ 1.8$ &      58 & $\mathrm{2s\,4f}$ & $\mathrm{^3F_{3  }^{o}}$ & $  2528570.$ & $  2530815.$ & $ 0.1$ & $  2526299.$ & $-0.1$ \\ 
    10 & $\mathrm{2p^2  }$ & $\mathrm{^1S_{0  }    }$ & $   553783.$ & $   567794.$ & $ 2.5$ & $   567267.$ & $ 2.4$ &      59 & $\mathrm{2s\,4f}$ & $\mathrm{^3F_{4  }^{o}}$ & $  2528570.$ & $  2530871.$ & $ 0.1$ & $  2526355.$ & $-0.1$ \\ 
    11 & $\mathrm{2s\,3s}$ & $\mathrm{^3S_{1  }    }$ & $  1855760.$ & $  1856089.$ & $ 0.0$ & $  1852844.$ & $-0.2$ &      60 & $\mathrm{2s\,4f}$ & $\mathrm{^1F_{3  }^{o}}$ & $        - $ & $  2533730.$ & $  - $ & $  2528991.$ & $  - $ \\ 
    12 & $\mathrm{2s\,3s}$ & $\mathrm{^1S_{0  }    }$ & $  1884420.$ & $  1886214.$ & $ 0.1$ & $  1882216.$ & $-0.1$ &      61 & $\mathrm{2p\,4s}$ & $\mathrm{^3P_{0  }^{o}}$ & $        - $ & $  2661563.$ & $  - $ & $  2655681.$ & $  - $ \\ 
    13 & $\mathrm{2s\,3p}$ & $\mathrm{^1P_{1  }^{o}}$ & $  1923850.$ & $  1925826.$ & $ 0.1$ & $  1922358.$ & $-0.1$ &      62 & $\mathrm{2p\,4s}$ & $\mathrm{^3P_{1  }^{o}}$ & $        - $ & $  2662724.$ & $  - $ & $  2656820.$ & $  - $ \\ 
    14 & $\mathrm{2s\,3p}$ & $\mathrm{^3P_{0  }^{o}}$ & $        - $ & $  1928630.$ & $  - $ & $  1925009.$ & $  - $ &      63 & $\mathrm{2p\,4s}$ & $\mathrm{^3P_{2  }^{o}}$ & $        - $ & $  2666840.$ & $  - $ & $  2661460.$ & $  - $ \\ 
    15 & $\mathrm{2s\,3p}$ & $\mathrm{^3P_{1  }^{o}}$ & $        - $ & $  1929220.$ & $  - $ & $  1925611.$ & $  - $ &      64 & $\mathrm{2p\,4s}$ & $\mathrm{^1P_{1  }^{o}}$ & $        - $ & $  2673641.$ & $  - $ & $  2666939.$ & $  - $ \\ 
    16 & $\mathrm{2s\,3p}$ & $\mathrm{^3P_{2  }^{o}}$ & $        - $ & $  1930059.$ & $  - $ & $  1926462.$ & $  - $ &      65 & $\mathrm{2p\,4p}$ & $\mathrm{^1P_{1  }    }$ & $        - $ & $  2681281.$ & $  - $ & $  2675347.$ & $  - $ \\ 
    17 & $\mathrm{2s\,3d}$ & $\mathrm{^3D_{1  }    }$ & $  1965860.$ & $  1967770.$ & $ 0.1$ & $  1964163.$ & $-0.1$ &      66 & $\mathrm{2p\,4p}$ & $\mathrm{^3D_{1  }    }$ & $        - $ & $  2684513.$ & $  - $ & $  2678722.$ & $  - $ \\ 
    18 & $\mathrm{2s\,3d}$ & $\mathrm{^3D_{2  }    }$ & $  1966080.$ & $  1967980.$ & $ 0.1$ & $  1964378.$ & $-0.1$ &      67 & $\mathrm{2p\,4p}$ & $\mathrm{^3D_{2  }    }$ & $        - $ & $  2684938.$ & $  - $ & $  2679129.$ & $  - $ \\ 
    19 & $\mathrm{2s\,3d}$ & $\mathrm{^3D_{3  }    }$ & $  1966300.$ & $  1968296.$ & $ 0.1$ & $  1964701.$ & $-0.1$ &      68 & $\mathrm{2p\,4p}$ & $\mathrm{^3D_{3  }    }$ & $        - $ & $  2688348.$ & $  - $ & $  2682876.$ & $  - $ \\ 
    20 & $\mathrm{2s\,3d}$ & $\mathrm{^1D_{2  }    }$ & $  1992340.$ & $  1997586.$ & $ 0.3$ & $  1993399.$ & $ 0.1$ &      69 & $\mathrm{2p\,4p}$ & $\mathrm{^3S_{1  }    }$ & $        - $ & $  2691166.$ & $  - $ & $  2684805.$ & $  - $ \\ 
    21 & $\mathrm{2p\,3s}$ & $\mathrm{^3P_{0  }^{o}}$ & $  2057140.$ & $  2055664.$ & $-0.1$ & $  2050249.$ & $-0.3$ &      70 & $\mathrm{2p\,4p}$ & $\mathrm{^3P_{0  }    }$ & $        - $ & $  2691420.$ & $  - $ & $  2685788.$ & $  - $ \\ 
    22 & $\mathrm{2p\,3s}$ & $\mathrm{^3P_{1  }^{o}}$ & $  2057140.$ & $  2057249.$ & $ 0.0$ & $  2051958.$ & $-0.3$ &      71 & $\mathrm{2p\,4p}$ & $\mathrm{^3P_{1  }    }$ & $        - $ & $  2694295.$ & $  - $ & $  2688609.$ & $  - $ \\ 
    23 & $\mathrm{2p\,3s}$ & $\mathrm{^3P_{2  }^{o}}$ & $  2057140.$ & $  2060910.$ & $ 0.2$ & $  2055983.$ & $-0.1$ &      72 & $\mathrm{2p\,4p}$ & $\mathrm{^3P_{2  }    }$ & $        - $ & $  2694733.$ & $  - $ & $  2689496.$ & $  - $ \\ 
    24 & $\mathrm{2p\,3s}$ & $\mathrm{^1P_{1  }^{o}}$ & $  2091870.$ & $  2090063.$ & $-0.1$ & $  2084057.$ & $-0.4$ &      73 & $\mathrm{2p\,4d}$ & $\mathrm{^3F_{2  }^{o}}$ & $        - $ & $  2697404.$ & $  - $ & $  2691501.$ & $  - $ \\ 
    25 & $\mathrm{2p\,3p}$ & $\mathrm{^1P_{1  }    }$ & $  2094820.$ & $  2097317.$ & $ 0.1$ & $  2093229.$ & $-0.1$ &      74 & $\mathrm{2p\,4d}$ & $\mathrm{^3F_{3  }^{o}}$ & $        - $ & $  2699888.$ & $  - $ & $  2694096.$ & $  - $ \\ 
    26 & $\mathrm{2p\,3p}$ & $\mathrm{^3D_{1  }    }$ & $  2102330.$ & $  2105510.$ & $ 0.2$ & $  2101249.$ & $-0.1$ &      75 & $\mathrm{2p\,4d}$ & $\mathrm{^1D_{2  }^{o}}$ & $        - $ & $  2700756.$ & $  - $ & $  2695172.$ & $  - $ \\ 
    27 & $\mathrm{2p\,3p}$ & $\mathrm{^3D_{2  }    }$ & $  2103900.$ & $  2107151.$ & $ 0.2$ & $  2102917.$ & $-0.0$ &      76 & $\mathrm{2p\,4p}$ & $\mathrm{^1D_{2  }    }$ & $  2696850.$ & $  2702087.$ & $ 0.2$ & $  2696558.$ & $-0.0$ \\ 
    28 & $\mathrm{2p\,3p}$ & $\mathrm{^3D_{3  }    }$ & $  2107390.$ & $  2110622.$ & $ 0.2$ & $  2106614.$ & $-0.0$ &      77 & $\mathrm{2p\,4d}$ & $\mathrm{^3F_{4  }^{o}}$ & $        - $ & $  2702915.$ & $  - $ & $  2697461.$ & $  - $ \\ 
    29 & $\mathrm{2p\,3p}$ & $\mathrm{^3S_{1  }    }$ & $  2119690.$ & $  2123472.$ & $ 0.2$ & $  2118706.$ & $-0.0$ &      78 & $\mathrm{2p\,4d}$ & $\mathrm{^3D_{1  }^{o}}$ & $        - $ & $  2705726.$ & $  - $ & $  2700008.$ & $  - $ \\ 
    30 & $\mathrm{2p\,3p}$ & $\mathrm{^3P_{0  }    }$ & $        - $ & $  2132252.$ & $  - $ & $  2126478.$ & $  - $ &      79 & $\mathrm{2p\,4d}$ & $\mathrm{^3D_{2  }^{o}}$ & $        - $ & $  2706671.$ & $  - $ & $  2701074.$ & $  - $ \\ 
    31 & $\mathrm{2p\,3p}$ & $\mathrm{^3P_{1  }    }$ & $  2128680.$ & $  2133785.$ & $ 0.2$ & $  2128247.$ & $-0.0$ &      80 & $\mathrm{2p\,4f}$ & $\mathrm{^1F_{3  }    }$ & $        - $ & $  2707963.$ & $  - $ & $  2701789.$ & $  - $ \\ 
    32 & $\mathrm{2p\,3p}$ & $\mathrm{^3P_{2  }    }$ & $  2130410.$ & $  2135720.$ & $ 0.2$ & $  2130292.$ & $-0.0$ &      81 & $\mathrm{2p\,4f}$ & $\mathrm{^3F_{3  }    }$ & $        - $ & $  2708388.$ & $  - $ & $  2702400.$ & $  - $ \\ 
    33 & $\mathrm{2p\,3d}$ & $\mathrm{^3F_{2  }^{o}}$ & $        - $ & $  2139901.$ & $  - $ & $  2135677.$ & $  - $ &      82 & $\mathrm{2p\,4d}$ & $\mathrm{^3D_{3  }^{o}}$ & $        - $ & $  2708430.$ & $  - $ & $  2703108.$ & $  - $ \\ 
    34 & $\mathrm{2p\,3d}$ & $\mathrm{^3F_{3  }^{o}}$ & $        - $ & $  2142598.$ & $  - $ & $  2138636.$ & $  - $ &      83 & $\mathrm{2p\,4f}$ & $\mathrm{^3F_{2  }    }$ & $        - $ & $  2708468.$ & $  - $ & $  2702641.$ & $  - $ \\ 
    35 & $\mathrm{2p\,3d}$ & $\mathrm{^1D_{2  }^{o}}$ & $  2141580.$ & $  2144437.$ & $ 0.1$ & $  2140459.$ & $-0.1$ &      84 & $\mathrm{2p\,4f}$ & $\mathrm{^3F_{4  }    }$ & $        - $ & $  2708819.$ & $  - $ & $  2702505.$ & $  - $ \\ 
    36 & $\mathrm{2p\,3d}$ & $\mathrm{^3F_{4  }^{o}}$ & $        - $ & $  2145366.$ & $  - $ & $  2141681.$ & $  - $ &      85 & $\mathrm{2p\,4d}$ & $\mathrm{^3P_{2  }^{o}}$ & $        - $ & $  2710818.$ & $  - $ & $  2705109.$ & $  - $ \\ 
    37 & $\mathrm{2p\,3p}$ & $\mathrm{^1D_{2  }    }$ & $  2148410.$ & $  2157057.$ & $ 0.4$ & $  2150745.$ & $ 0.1$ &      86 & $\mathrm{2p\,4d}$ & $\mathrm{^3P_{1  }^{o}}$ & $        - $ & $  2711544.$ & $  - $ & $  2705747.$ & $  - $ \\ 
    38 & $\mathrm{2p\,3d}$ & $\mathrm{^3D_{1  }^{o}}$ & $  2160650.$ & $  2165232.$ & $ 0.2$ & $  2159945.$ & $-0.0$ &      87 & $\mathrm{2p\,4d}$ & $\mathrm{^3P_{0  }^{o}}$ & $        - $ & $  2711943.$ & $  - $ & $  2706094.$ & $  - $ \\ 
    39 & $\mathrm{2p\,3d}$ & $\mathrm{^3D_{2  }^{o}}$ & $  2161960.$ & $  2165961.$ & $ 0.2$ & $  2160762.$ & $-0.1$ &      88 & $\mathrm{2p\,4f}$ & $\mathrm{^3P_{3  }    }$ & $        - $ & $  2712961.$ & $  - $ & $  2706897.$ & $  - $ \\ 
    40 & $\mathrm{2p\,3d}$ & $\mathrm{^3D_{3  }^{o}}$ & $  2163340.$ & $  2167469.$ & $ 0.2$ & $  2162400.$ & $-0.0$ &      89 & $\mathrm{2p\,4f}$ & $\mathrm{^3P_{4  }    }$ & $        - $ & $  2713531.$ & $  - $ & $  2707350.$ & $  - $ \\ 
    41 & $\mathrm{2p\,3d}$ & $\mathrm{^3P_{2  }^{o}}$ & $  2170190.$ & $  2174402.$ & $ 0.2$ & $  2169739.$ & $-0.0$ &      90 & $\mathrm{2p\,4f}$ & $\mathrm{^3P_{5  }    }$ & $        - $ & $  2715225.$ & $  - $ & $  2708632.$ & $  - $ \\ 
    42 & $\mathrm{2p\,3d}$ & $\mathrm{^3P_{1  }^{o}}$ & $  2171680.$ & $  2175473.$ & $ 0.2$ & $  2170906.$ & $-0.0$ &      91 & $\mathrm{2p\,4f}$ & $\mathrm{^3D_{3  }    }$ & $        - $ & $  2716473.$ & $  - $ & $  2710743.$ & $  - $ \\ 
    43 & $\mathrm{2p\,3d}$ & $\mathrm{^3P_{0  }^{o}}$ & $        - $ & $  2176019.$ & $  - $ & $  2171512.$ & $  - $ &      92 & $\mathrm{2p\,4f}$ & $\mathrm{^1D_{4  }    }$ & $        - $ & $  2717105.$ & $  - $ & $  2710218.$ & $  - $ \\ 
    44 & $\mathrm{2p\,3p}$ & $\mathrm{^1S_{0  }    }$ & $        - $ & $  2193603.$ & $  - $ & $  2186094.$ & $  - $ &      93 & $\mathrm{2p\,4f}$ & $\mathrm{^3D_{2  }    }$ & $        - $ & $  2717141.$ & $  - $ & $  2711361.$ & $  - $ \\ 
    45 & $\mathrm{2p\,3d}$ & $\mathrm{^1F_{3  }^{o}}$ & $  2192860.$ & $  2203326.$ & $ 0.5$ & $  2195894.$ & $ 0.1$ &      94 & $\mathrm{2p\,4p}$ & $\mathrm{^1S_{0  }    }$ & $        - $ & $  2717922.$ & $  - $ & $  2706869.$ & $  - $ \\ 
    46 & $\mathrm{2p\,3d}$ & $\mathrm{^1P_{1  }^{o}}$ & $        - $ & $  2208387.$ & $  - $ & $  2201414.$ & $  - $ &      95 & $\mathrm{2p\,4f}$ & $\mathrm{^3D_{1  }    }$ & $        - $ & $  2718397.$ & $  - $ & $  2712736.$ & $  - $ \\ 
    47 & $\mathrm{2s\,4s}$ & $\mathrm{^3S_{1  }    }$ & $        - $ & $  2477026.$ & $  - $ & $  2472580.$ & $  - $ &      96 & $\mathrm{2p\,4f}$ & $\mathrm{^1D_{2  }    }$ & $        - $ & $  2719895.$ & $  - $ & $  2713991.$ & $  - $ \\ 
    48 & $\mathrm{2s\,4s}$ & $\mathrm{^1S_{0  }    }$ & $        - $ & $  2488107.$ & $  - $ & $  2483578.$ & $  - $ &      97 & $\mathrm{2p\,4d}$ & $\mathrm{^1F_{3  }^{o}}$ & $        - $ & $  2722212.$ & $  - $ & $  2715080.$ & $  - $ \\ 
    49 & $\mathrm{2s\,4p}$ & $\mathrm{^3P_{0  }^{o}}$ & $        - $ & $  2504555.$ & $  - $ & $  2500383.$ & $  - $ &      98 & $\mathrm{2p\,4d}$ & $\mathrm{^1P_{1  }^{o}}$ & $        - $ & $  2723448.$ & $  - $ & $  2716711.$ & $  - $ \\ 
   \hline
\end{tabular}
\end{small}
\end{center}
\end{minipage}
\end{table*}

%% file: belike2.bbl
\begin{thebibliography}{}

\bibitem[\protect\citeauthoryear{Aggarwal \& Keenan}{Aggarwal \&
  Keenan}{2012}]{aggarwal2012a}
Aggarwal K.~M.,  Keenan F.~P.,  2012, Phys. Scr., 86, 055301

\bibitem[\protect\citeauthoryear{Aggarwal \& Keenan}{Aggarwal \&
  Keenan}{2014a}]{aggarwal2014a}
Aggarwal K.~M.,  Keenan F.~P.,  2014a, Phys. Scr., 89, 125401

\bibitem[\protect\citeauthoryear{Aggarwal \& Keenan}{Aggarwal \&
  Keenan}{2014b}]{aggarwal2014c}
Aggarwal K.~M.,  Keenan F.~P.,  2014b, Mon. Not. R. Astr. Soc., 445, 2015

\bibitem[\protect\citeauthoryear{Aggarwal \& Keenan}{Aggarwal \&
  Keenan}{2014c}]{aggarwal2014b}
Aggarwal K.~M.,  Keenan F.~P.,  2014c, Mon. Not. R. Astr. Soc., 438, 1223

\bibitem[\protect\citeauthoryear{Aggarwal \& Keenan}{Aggarwal \&
  Keenan}{2015a}]{aggarwal2015}
Aggarwal K.~M.,  Keenan F.~P.,  2015a, Mon. Not. R. Astr. Soc., 447, 3849

\bibitem[\protect\citeauthoryear{Aggarwal \& Keenan}{Aggarwal \&
  Keenan}{2015b}]{aggarwal2015b}
Aggarwal K.~M.,  Keenan F.~P.,  2015b, Mon. Not. R. Astr. Soc., 0, submitted
  http://arxiv.org/abs/1503.07673

\bibitem[\protect\citeauthoryear{Badnell}{Badnell}{1999}]{badnell1999c}
Badnell N.~R.,  1999, Journal of Physics B: Atomic, Molecular and Optical
  Physics, 32, 5583

\bibitem[\protect\citeauthoryear{Badnell}{Badnell}{2011}]{badnell2011b}
Badnell N.~R.,  2011, Comput. Phys. Commun., 182, 1528

\bibitem[\protect\citeauthoryear{Badnell \& Ballance}{Badnell \&
  Ballance}{2014}]{badnell2014}
Badnell N.~R.,  Ballance C.~P.,  2014, The Astrophysical Journal, 785, 99

\bibitem[\protect\citeauthoryear{Badnell \& Griffin}{Badnell \&
  Griffin}{1999}]{badnell1999a}
Badnell N.~R.,  Griffin D.~C.,  1999, J. Phys. B, 32, 2267

\bibitem[\protect\citeauthoryear{Berrington, Ballance, Griffin \&
  Badnell}{Berrington et~al.}{2005}]{berrington2005}
Berrington K.~A.,  Ballance C.~P.,  Griffin D.~C.,    Badnell N.~R.,  2005,
  Journal of Physics B: Atomic, Molecular and Optical Physics, 38, 1667

\bibitem[\protect\citeauthoryear{Berrington, Burke, Butler, Seaton, Storey,
  Taylor \& Yan}{Berrington et~al.}{1987}]{berrington1987}
Berrington K.~A.,  Burke P.~G.,  Butler K.,  Seaton M.~J.,  Storey P.~J.,
  Taylor K.~T.,    Yan Y.,  1987, Journal of Physics B: Atomic and Molecular
  Physics, 20, 6379

\bibitem[\protect\citeauthoryear{Berrington, Eissner \& Norrington}{Berrington
  et~al.}{1995}]{berrington1995}
Berrington K.~A.,  Eissner W.~B.,    Norrington P.~H.,  1995, Comput. Phys.
  Commun., 92, 290

\bibitem[\protect\citeauthoryear{Bryans, Badnell, Gorczyca, Laming, Mitthumsiri
  \& Savin}{Bryans et~al.}{2006}]{bryans2006}
Bryans P.,  Badnell N.~R.,  Gorczyca T.~W.,  Laming J.~M.,  Mitthumsiri W.,
  Savin D.~W.,  2006, Astrophys. J. Suppl. Ser., 167, 343

\bibitem[\protect\citeauthoryear{Burgess \& Tully}{Burgess \&
  Tully}{1992}]{burgess1992}
Burgess A.,  Tully J.~A.,  1992, Astron. Astrophys., 254, 436

\bibitem[\protect\citeauthoryear{Fern\'andez-Menchero, {Del~Zanna} \&
  Badnell}{Fern\'andez-Menchero et~al.}{2014a}]{fernandez-menchero2014a}
Fern\'andez-Menchero L.,  {Del~Zanna} G.,    Badnell N.~R.,  2014a, Astron.
  Astrophys., 566, A104

\bibitem[\protect\citeauthoryear{Fern\'andez-Menchero, {Del~Zanna} \&
  Badnell}{Fern\'andez-Menchero et~al.}{2014b}]{fernandez-menchero2014b}
Fern\'andez-Menchero L.,  {Del~Zanna} G.,    Badnell N.~R.,  2014b, Astron.
  Astrophys., 572, A115

\bibitem[\protect\citeauthoryear{Griffin, Badnell \& Pindzola}{Griffin
  et~al.}{1998}]{griffin1998}
Griffin D.~C.,  Badnell N.~R.,    Pindzola M.~S.,  1998, J. Phys. B, 31, 3713

\bibitem[\protect\citeauthoryear{Griffin, Badnell, Pindzola \& Shaw}{Griffin
  et~al.}{1999}]{griffin1999}
Griffin D.~C.,  Badnell N.~R.,  Pindzola M.~S.,    Shaw J.~A.,  1999, J. Phys.
  B, 32, 2139

\bibitem[\protect\citeauthoryear{Hummer, Berrington, Eissner, Pradhan, Saraph
  \& Tully}{Hummer et~al.}{1993}]{hummer1993}
Hummer D.~G.,  Berrington K.~A.,  Eissner W.,  Pradhan A.~K.,  Saraph H.~E.,
  Tully J.~A.,  1993, Astron. Astrophys., 279, 298

\bibitem[\protect\citeauthoryear{Kallman \& Bautista}{Kallman \&
  Bautista}{2001}]{kallman2001}
Kallman T.,  Bautista M.,  2001, The Astrophysical Journal Supplement Series,
  133, 221

\bibitem[\protect\citeauthoryear{Keenan}{Keenan}{1988}]{keenan1988b}
Keenan F.~P.,  1988, Physica Scripta, 37, 57

\bibitem[\protect\citeauthoryear{Liang \& Badnell}{Liang \&
  Badnell}{2010}]{liang2010a}
Liang G.~Y.,  Badnell N.~R.,  2010, Astron. Astrophys., 518, A64

\bibitem[\protect\citeauthoryear{Liang, Badnell, L\'opez-Urrutia, Baumann,
  {Del~Zanna}, Storey, Tawara \& Ullrich}{Liang et~al.}{2010}]{liang2010b}
Liang G.~Y.,  Badnell N.~R.,  L\'opez-Urrutia J.~R.~C.,  Baumann T.~M.,
  {Del~Zanna} G.,  Storey P.~J.,  Tawara H.,    Ullrich J.,  2010, The
  Astrophysical Journal Supplement Series, 190, 322

\bibitem[\protect\citeauthoryear{Liang, Badnell \& Zhao}{Liang
  et~al.}{2012}]{liang2012}
Liang G.~Y.,  Badnell N.~R.,    Zhao G.,  2012, Astron. Astrophys., 547, A87

\bibitem[\protect\citeauthoryear{Liang, Whiteford \& Badnell}{Liang
  et~al.}{2009}]{liang2009b}
Liang G.~Y.,  Whiteford A.~D.,    Badnell N.~R.,  2009, Astron. Astrophys.,
  500, 1263

\bibitem[\protect\citeauthoryear{Martin \& Zalubas}{Martin \&
  Zalubas}{1979}]{martin1979}
Martin W.~C.,  Zalubas R.,  1979, J. Phys. Chem. Ref. Data, 8, 817

\bibitem[\protect\citeauthoryear{Storey, Sochi \& Badnell}{Storey
  et~al.}{2014}]{storey2014b}
Storey P.~J.,  Sochi T.,    Badnell N.~R.,  2014, Monthly Notices of the Royal
  Astronomical Society, 441, 3028

\end{thebibliography}
